\documentclass[aps,prl,reprint,showpacs,amsmath,amssymb,floatfix,superscriptaddress,noeprint]{revtex4-2}

\usepackage[colorlinks,linkcolor=blue,urlcolor=blue,citecolor=blue,anchorcolor=blue]{hyperref}

\usepackage{graphicx} 

\usepackage{physics}
\usepackage{siunitx}
\DeclareSIUnit{\belmilliwatt}{Bm}
\DeclareSIUnit{\dBm}{\deci\belmilliwatt}
\usepackage{tensor}

\usepackage{xcolor}

\usepackage{stackengine}
\stackMath
\newcommand\subline[2]{\ensuremath{\stackon[-1.5pt]{#1}{\rule[2pt]{\widthof{$#1$}}{.4pt}_{#2}}}}

\newcommand{\sigz}{\ensuremath{\hat{\sigma}_z}}
\newcommand{\sigy}{\ensuremath{\hat{\sigma}_y}}
\newcommand{\sigx}{\ensuremath{\hat{\sigma}_x}}
\newcommand{\sigp}{\ensuremath{\hat{\sigma}_+}}
\newcommand{\sigm}{\ensuremath{\hat{\sigma}_-}}

\renewcommand{\wr}{\ensuremath{\omega_{\mathrm{R}}}}
\newcommand{\wq}{\ensuremath{\omega_{\mathrm{Q}}}}

\makeatletter
\newsavebox\myboxA
\newsavebox\myboxB
\newlength\mylenA

\newcommand*\xoverline[2][0.9]{%
    \sbox{\myboxA}{$\m@th#2$}%
    \setbox\myboxB\null
    \ht\myboxB=\ht\myboxA%
    \dp\myboxB=\dp\myboxA%
    \wd\myboxB=#1\wd\myboxA
    \sbox\myboxB{$\m@th\overline{\copy\myboxB}$}
    \setlength\mylenA{\the\wd\myboxA}
    \addtolength\mylenA{-\the\wd\myboxB}%
    \ifdim\wd\myboxB<\wd\myboxA%
       \rlap{\hskip 0.5\mylenA\usebox\myboxB}{\usebox\myboxA}%
    \else
        \hskip -0.5\mylenA\rlap{\usebox\myboxA}{\hskip 0.5\mylenA\usebox\myboxB}%
    \fi}
\makeatother

\begin{document}
\renewcommand\stackalignment{l}
\title{Energetics of a Single Qubit Gate}


\author{J. Stevens}
\thanks{These authors have contributed equally.}
\affiliation{Ecole Normale Sup\'erieure de Lyon,  CNRS, Laboratoire de Physique, F-69342 Lyon, France}
\author{D. Szombati}
\thanks{These authors have contributed equally.}
\affiliation{Ecole Normale Sup\'erieure de Lyon,  CNRS, Laboratoire de Physique, F-69342 Lyon, France}
\author{M. Maffei}
\affiliation{CNRS and Universit\'e Grenoble Alpes, Institut N\'eel, F-38042 Grenoble, France}
\author{C. Elouard}
\affiliation{QUANTIC team, INRIA de Paris, 2 Rue Simone Iff, 75012 Paris, France}
\author{R. Assouly}
\affiliation{Ecole Normale Sup\'erieure de Lyon,  CNRS, Laboratoire de Physique, F-69342 Lyon, France}
\author{N. Cottet}
\affiliation{Ecole Normale Sup\'erieure de Lyon,  CNRS, Laboratoire de Physique, F-69342 Lyon, France}
\author{R. Dassonneville}
\affiliation{Ecole Normale Sup\'erieure de Lyon,  CNRS, Laboratoire de Physique, F-69342 Lyon, France}
\author{Q. Ficheux}
\affiliation{Ecole Normale Sup\'erieure de Lyon,  CNRS, Laboratoire de Physique, F-69342 Lyon, France}
\author{S. Zeppetzauer}
\affiliation{Ecole Normale Sup\'erieure de Lyon,  CNRS, Laboratoire de Physique, F-69342 Lyon, France}
\author{A. Bienfait}
\affiliation{Ecole Normale Sup\'erieure de Lyon,  CNRS, Laboratoire de Physique, F-69342 Lyon, France}
\author{A. N. Jordan}
\affiliation{Institute for Quantum Studies, Chapman University, 1 University Drive, Orange, CA 92866, USA}
\affiliation{Department of Physics and Astronomy, University of Rochester, Rochester, New York 14627, USA}
\author{A. Auff\`eves}
\affiliation{CNRS and Universit\'e Grenoble Alpes, Institut N\'eel, F-38042 Grenoble, France}
\author{B. Huard}
\affiliation{Ecole Normale Sup\'erieure de Lyon,  CNRS, Laboratoire de Physique, F-69342 Lyon, France}
\date{\today}
\begin{abstract}
Qubits are physical, a quantum gate thus not only acts on the information carried by the qubit but also on its energy. What is then the corresponding flow of energy between the qubit and the controller that implements the gate? Here we exploit a superconducting platform to answer this question in the case of a quantum gate realized by a resonant drive field. During the gate, the superconducting qubit becomes entangled with the microwave drive pulse so that there is a quantum superposition between energy flows. We measure the energy change in the drive field conditioned on the outcome of a projective qubit measurement. We demonstrate that the drive's energy change associated with the measurement backaction can exceed by far the energy that can be extracted by the qubit. This can be understood by considering the qubit as a weak measurement apparatus of the driving field.
\end{abstract}
\maketitle

Understanding the energetic resources needed to operate quantum computers is crucial to assess their performance limitations~\cite{Barnes1999,gea-banacloche_implications_2002,gea-banacloche_minimum_2002,ozawa_conservative_2002,Karasawa2007,Bedingham16,ikonen_energy-efficient_2017,Guryanova20,Cimini2020,Deffner2021}. Beyond the fundamental costs associated with information processing~\cite{Faist15}, e.g. reset~\cite{Reeb14} and measurements~\cite{Sagawa09,Mohammady19}, quantum gates need energy to manipulate qubits encoded in non-degenerate states~\cite{Deffner2021,Aifer2022}.
Since a gate can prepare a quantum superposition of states with different energies, the energy balance between the gate controller and the qubit can be seen as a quantum superposition of energetic costs. Focusing on gates performed by resonant driving, the drive appears to have exchanged energy with the qubit. Yet the amount of transferred energy is undetermined until the qubit state is measured. How is the energy in the driving mode modified by the qubit measurement and what does it reveal about the qubit-drive system? Superconducting circuits offer a state-of-the-art platform for exploring this question owing to the possibility to perform single shot qubit readout using an ancillary cavity and quantum-limited measurements of propagating microwave modes \cite{gu_microwave_2017}. In particular, it is possible to manipulate~\cite{houck_generating_2007,hoi_demonstration_2011,hoi_generation_2012,reuer_realization_2021} and probe~\cite{astafiev_resonance_2010,abdumalikov_dynamics_2011,honigl-decrinis_two-level_2020,lu_characterizing_2021} the fields interacting resonantly with the qubit. Superconducting circuits have thus been useful to explore quantum thermodynamics properties of their spontaneous or stimulated emission~\cite{cottet_observing_2017,naghiloo_heat_2020,binder_thermodynamics_2018,Lu2021}, and build quantum thermal engines~\cite{ronzani_tunable_2018,senior_heat_2020,Lu2022}.
Correlations between the resonant drive amplitude and the outcome of a later qubit measurement have been evidenced by probing quantum trajectories of superconducting qubits~\cite{campagne-ibarcq_observing_2016,jordan_anatomy_2016,naghiloo_mapping_2016,ficheux_dynamics_2018,Scigliuzzo2020} including when a projective measurement is used to perform post-selection~\cite{campagne-ibarcq_observing_2014,tan_homodyne_2017,naghiloo_quantum_2017}. However the demonstration of correlations between the energy of the drive mode and the qubit state is missing.

In this Letter, we present an experiment in which we directly probe the energy in the driving mode conditioned on the measured qubit state. 
 We observe that measuring the qubit energy leads to a change in the  energy of the driving pulse owing to its entanglement with the qubit before measurement. Strikingly, we also observe that the energy of the pulse can change by more than a quantum depending on the measured qubit state, revealing a subtle backaction of the qubit measurement on the drive pulse.

\begin{figure}
\centering
\includegraphics[width=\linewidth]{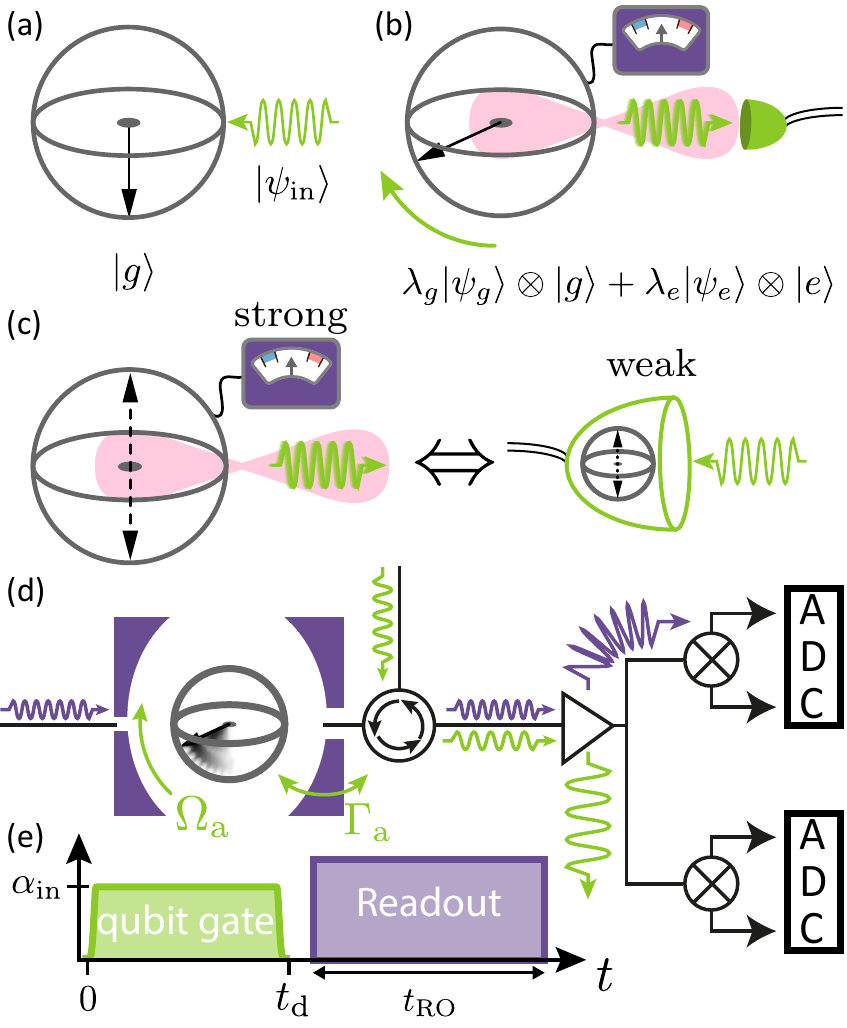}
\caption{Principle of the experiment. a. Coherent wavepacket $\ket{\psi_\text{in}}$ (green arrow) at the qubit frequency interacts with a qubit prepared in $\ket{g}$ (Bloch vector). b. Resulting entangled state Eq.~(\ref{eq:bipartite}). The energy of the outgoing drive wavepacket is measured and averaged conditionally on the outcome of a strong readout of the qubit energy.
c. Schematics highlighting the equivalence between the action of the projective qubit measurement and that of a weak measurement apparatus on the pulse.
d. The transmon qubit is placed inside a microwave cavity (purple) to perform its readout by sending a pulse at the cavity frequency through a weakly coupled port (left). The resonant field (green) addressing the qubit is sent through a strongly coupled port on the right. Both pulses exit through this port and are directed by a circulator into low noise amplifiers. Their quadratures are measured via two heterodyne setups based on Analog-to-Digital Converters (ADCs) operating at qubit and cavity frequencies. e. Scheme of the experimental pulse sequence, where $t_\mathrm{d}=400~\mathrm{ns}$ and $t_{\mathrm{RO}} =\SI{704}{\nano\second}$~\cite{noauthor_supplementary_nodate}.}
\label{fig:figure_1}
\end{figure}

In order to better understand the rise of these correlations, let us consider the joint evolution of the qubit and drive mode during the qubit gate. 
Assuming the qubit starts in the ground state $\ket{g}$, and is driven by a coherent state $\ket{\psi_\text{in}}$, the qubit and the propagating drive mode $a$ are initially in the separable state 
$\ket{\psi_\text{in}}\otimes\ket{g}$ (see Fig. \ref{fig:figure_1}a). 
Owing to the light-matter coupling between the drive mode and the qubit, they evolve into the entangled state \cite{gea-banacloche_minimum_2002,vanEnk2002,Silberfarb2004}
\begin{equation}
\lambda_g\ket{\psi_g}\otimes\ket{g} + \lambda_e\ket{\psi_e}\otimes\ket{e} \, \label{eq:bipartite}
\end{equation}
where $\lambda_g$ and $\lambda_e$ are the probability amplitudes for each state in the superposition, and $\ket{\psi_{g,e}}$ designate the outgoing states of the drive mode (see Fig. \ref{fig:figure_1}b). Note that these parameters and states depend on $|\psi_\mathrm{in}\rangle$ implicitly. The qubit gate is parametrized by the rotation of angle $\theta$ undergone by the qubit Bloch vector, revealed by tracing over the field. 
Interestingly, the  entanglement above limits the fidelity of a qubit gate, a question which has been at the core of an intense two decades old debate~\cite{vanEnk2002,gea-banacloche_implications_2002,Itano2003,gea-banacloche_minimum_2002,ozawa_conservative_2002,Silberfarb2003,Nha2005,Gea_Banacloche_2005,ikonen_energy-efficient_2017}, since the purity of the qubit density matrix $\rho$ reads 
\begin{equation}
\mathrm{tr}(\rho^2)=1-2|\lambda_g\lambda_e|^2\left(1-|\langle \psi_e|\psi_g\rangle|^2\right).
\end{equation}
Luckily for quantum computing, it is possible to reach large gate fidelity since the minimum gate error $1-\mathrm{tr}(\rho^2)$ scales as the inverse of the average photon number in $|\psi_\mathrm{in}\rangle$~\cite{Barnes1999,gea-banacloche_minimum_2002,bertet_complementarity_2001,vanEnk2002}. The lack of purity also determines how much information can be extracted about the drive mode when measuring the qubit state. When the qubit is measured, the measurement backaction prepares the drive mode $a$ in states of different energy expectations. Conservation of the expected energy before and after the resonant interaction leads to the following equality relating the expected number of quanta in the initial state $\ket{\psi_\text{in}}\otimes\ket{g}$ and the final state (\ref{eq:bipartite})
\begin{equation}
\langle a^\dagger a\rangle_{|\psi_{\mathrm{in}}\rangle}=
|\lambda_g|^2\langle a^\dagger a\rangle_{|\psi_g\rangle}+|\lambda_e|^2\left[\langle a^\dagger a\rangle_{|\psi_e\rangle}+1\right].
\end{equation}
In this work, we directly measure the energy contained in the states  $\ket{\psi_g}$ and $\ket{\psi_e}$, and its dependence on the drive amplitude. Interestingly, from the point of view of the driving mode, the qubit acts as a weak measurement apparatus, which exerts a backaction that our experiment is able to probe (Fig.~\ref{fig:figure_1}c).

Our setup is schematically represented in Fig.~\ref{fig:figure_1}d~\cite{noauthor_supplementary_nodate}. A transmon qubit of frequency $\wq = 2 \pi \times \SI{4.81}{\giga\hertz}$  is embedded in an superconducting cavity of frequency $\wr = 2 \pi \times \SI{7.69}{\giga\hertz}$ below $15~\mathrm{mK}$. The qubit relaxation time $T_1=5.5 \pm 0.3~\si{\micro\second}$ is mainly limited by its coupling rate $\Gamma_a=2\pi\times 20~\mathrm{kHz}$ to a transmission line that carries the driving mode $a$. The qubit  pure dephasing time is $T_\varphi=2.4~\mu\mathrm{s}$. 

We perform the following experiment. First, a pulse of varying amplitude $\alpha_\text{in}$, whose phase is chosen so that $\alpha_\text{in}>0$, drives the qubit at frequency $\wq$ for a fixed duration $t_\mathrm{d} = \SI{400}{\nano\second}$ (Fig. \ref{fig:figure_1}e). The pulse is reflected and amplified using a Travelling Wave Parametric Amplifier (TWPA)~\cite{macklin_nearquantum-limited_2015}. A heterodyne measurement yields a continuous record of its two quadratures. This drive pulse induces a rotation of the qubit of angle $\theta$ around $\sigma_y$. The qubit is then measured dispersively 20~ns later using a $\SI{704}{\nano\second}$-long pulse at the cavity frequency $\wr$ sent on a weakly-coupled auxiliary port. This readout pulse exits through the strongly coupled output port used for driving the qubit and its transmission is detected through the same amplification chain.

\begin{figure}
\centering
\includegraphics[width=\linewidth]{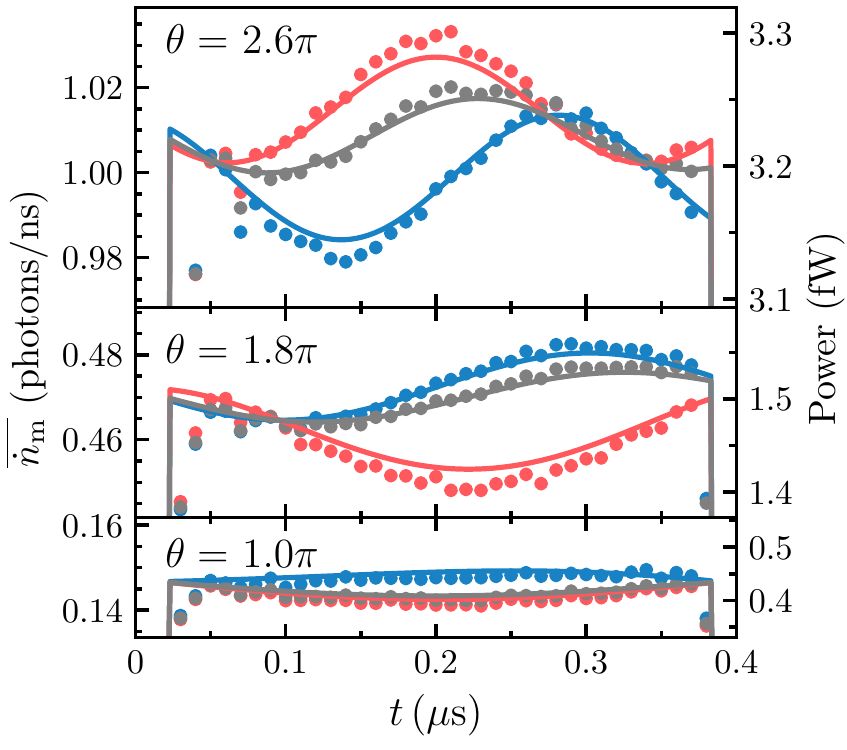}
\caption{Measured power of the reflected drive. Dots: mean instantaneous power  $\subline{\dot{n}_\mathrm{m}}{}$ of the outgoing drive in units of photon flux~\cite{noauthor_supplementary_nodate} as a function of time $t$. Each panel corresponds to a different input drive amplitude resulting in qubit rotation angles $\theta = \pi$, $1.8\pi$ and $2.6\pi$. Gray: averaging without post-selection, blue (red): averaging conditioned on the qubit being measured in $|g\rangle$ ($|e\rangle$). Lines: expected power from Eq.~(\ref{eq:weakpow}). The time delay between the experimental and numerical data has been adjusted by hand.}
\label{fig:figure_2}
\end{figure}

We  start  by  measuring  the  average  energy  in  the reflected drive pulse. From the heterodyne measurement it is possible to access both the complex amplitude $\alpha_\mathrm{m}$ and the instantaneous outgoing power $\dot{n}_\mathrm{m}$  (in units of photons per second) referred to the qubit output port~\cite{noauthor_supplementary_nodate}. To account for the added noise of the amplifiers and possible experimental gain drifts, we interleave the measurement with a calibration sequence where the qubit is shifted out of resonance using the ac-Stark effect. The average measured photon flux outgoing from the qubit in state $\rho$ is given by~ \cite{vool_introduction_2017,clerk_introduction_2010,cottet_energy_2018}
\begin{equation}
    \subline{\dot{n}_\mathrm{m}}{\rho} = \alpha_\mathrm{in}^2 - \frac{\Omega_a}{2}\expval{\sigx}_\rho + \Gamma_a \frac{1 + \expval{\sigz}_\rho}{2}  \label{eq:expval_pow}
\end{equation}
where $\Omega_a= 2 \sqrt{\Gamma_a} \alpha_\text{in}$ denotes the Rabi frequency and $\hat{\sigma}_{x,y,z}$ are the three Pauli matrices. In Fig. \ref{fig:figure_2}a, we show the evolution of $\subline{\dot{n}_\mathrm{m}}{\rho}$ for varying input drive powers. This temporal version of the Mollow triplet was already observed in several experiments~\cite{astafiev_resonance_2010,abdumalikov_dynamics_2011,honigl-decrinis_two-level_2020,lu_characterizing_2021}. 

To extract the correlation between the power of the re-emitted microwave drive and final qubit state, we average the instantaneous power conditioned on the measured qubit state (Fig.~\ref{fig:figure_2}). We observe that a clear deviation exists from the unconditional average power. Theoretically, it is possible to capture the dependence of the drive power on qubit measurement outcome using the past quantum state formalism \cite{wiseman_weak_2002,tsang_optimal_2009,gammelmark_past_2013}. A full description of the drive mode at each moment in time can be given by considering both the initial starting condition via the density matrix of the qubit $\rho(t)$ and the final measurement result through the effect matrix of the qubit $E(t)$. 
The density matrix obeys the standard Lindblad equation while the effect matrix is constrained by its value at the final measurement time and is back propagated using the adjoint of the Lindblad equation (see~\cite{noauthor_supplementary_nodate}). This formalism was used in Ref.~\cite{campagne-ibarcq_observing_2014} in order to determine the post-selected average evolution of the transmitted drive amplitude through a qubit. For a reflected drive, the post-selected average measured drive amplitude reads
\begin{equation}
\subline{\alpha_\mathrm{m}}{E,\rho} = \alpha_\text{in} - \sqrt{\Gamma_a} \ \Re[\tensor*[_E]{\expval{\sigm}}{_\rho}] ,\label{eq:weakampl}
\end{equation}
where $\tensor*[_E]{\expval{\sigm}}{_\rho} = \frac{\Tr\left[E(t)\sigm \rho(t) \right]}{\Tr\left[E(t)\rho(t)\right]}$ is the weak value of the qubit lowering operator $\sigm = (\sigx - i \sigy)/2$~ \cite{campagne-ibarcq_observing_2014}. The coherent part of the power emitted by the qubit corresponds to the modulus square of that amplitude. In contrast, in this work we are concerned with the total energy contained in the drive mode, and not only the coherent part. One can show that the post-selected expectation value of the outgoing photon flux is given by~\cite{maffei_probing_2021,maffei_manuscript_nodate}

\begin{equation}
\subline{\dot{n}_\mathrm{m}}{E,\rho} =
\abs{\alpha_\text{in}}^2 - \Omega_a \Re[\tensor*[_E]{\expval{\sigm}}{_\rho}]+ \Gamma_a \frac{\Tr\left[E \sigm \rho \sigp \right]}{\Tr\left[E \rho \right]} \ ,\label{eq:weakpow}
\end{equation}
where the last term is the weak value of a photo-detection rate. 
To compute Eq.~(\ref{eq:weakpow}), we solve the forward and backward Lindblad equations. An independent measurement allows us to set $\rho(0)$ to a thermal state with an excitation probability $0.088\pm0.002$. The effect matrix $E$ is set at measurement time $t = t_\mathrm{d}$ conditionally on the post-selected readout outcome. When the qubit is measured in state $|e\rangle$ with a readout fidelity $F_e = 0.867 \pm 0.028$, it is given by $E_e(t_\mathrm{d}) = F_e\ket{e}\bra{e}+(1-F_e)\ket{g}\bra{g}$, while, when the qubit is measured in state $|g\rangle$ with a readout fidelity $F_g=0.985\pm 0.015$, it is $E_g(t_\mathrm{d}) = F_g\ket{g}\bra{g}+(1-F_g)\ket{e}\bra{e}$~\cite{noauthor_supplementary_nodate}. Note that without post-selection, the effect matrix is the identity and Eq.~(\ref{eq:weakpow}) comes down to the non post-selected case in Eq.~(\ref{eq:expval_pow}).
The Eq.~(\ref{eq:weakpow}) reproduces the measured post-selected instantaneous powers we observe (solid lines Fig. \ref{fig:figure_2}), where the single fit parameter is the electrical delay of the setup.

\begin{figure}
\centering
\includegraphics[width=\linewidth]{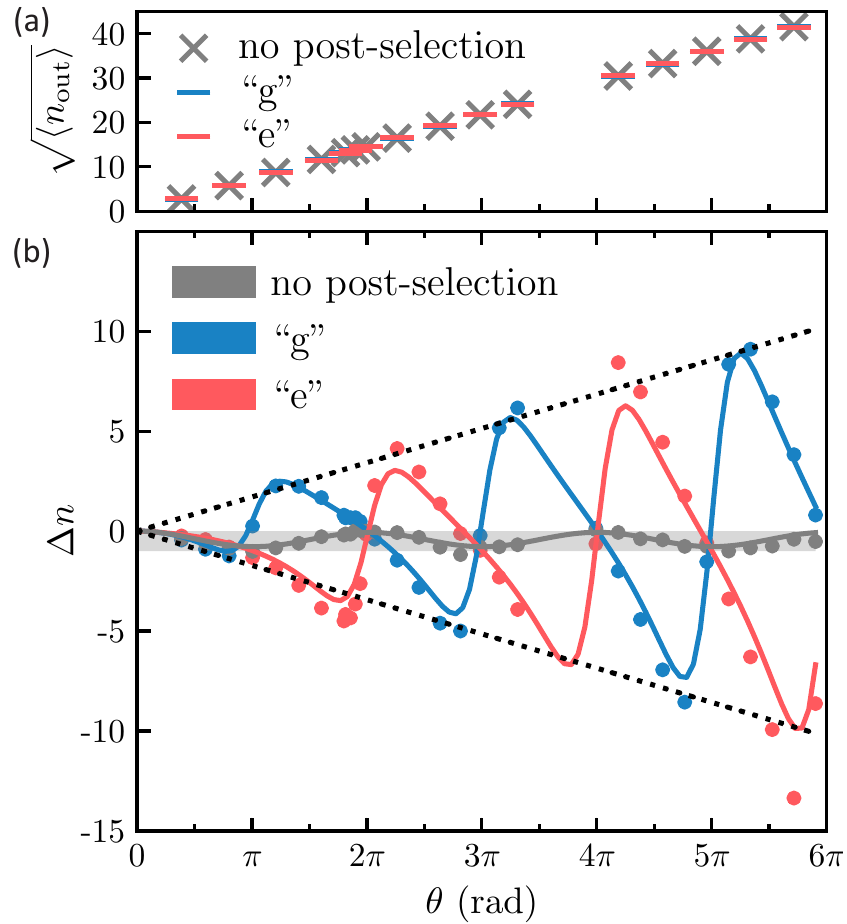}
\caption{a. Square root of the measured total mean number of photons in the outgoing drive pulse as a function of the qubit rotation angle $\theta$ around $\sigy$ for post-selected and non post-selected data. For these photon numbers, the effect of post-selection is almost indistinguishable. b. Dots: measured difference $\Delta n$ between the mean post-selected number of photons and the mean number of photons in the incoming drive pulse as a function of the qubit rotation angle. Colors indicate the kind of post-selection. Lines: time integrated Eq.~(\ref{eq:weakpow}). Dotted lines: guides to the eye scaling with $\theta\propto\sqrt{n_\mathrm{in}}$. Shaded area: allowed range of exchanged energy without post-selection (between $-1$ and $0$ photons).}
\label{fig:figure_3}
\end{figure}

Our original motivation is to quantify the difference of energy between the post-selected drive pulses. The total number of photons contained in the pulse can be calculated as $\expval{n_\mathrm{out}} = \int_0^{t_\mathrm{d}} \subline{\dot{n}_\mathrm{m}}{E,\rho} \dd t$ from the experimental data. In Fig.~\ref{fig:figure_3}a, we show the square root of the measured total photon numbers $\sqrt{\expval{n_\mathrm{out}}}$ as a function of the rotation angle $\theta$ in the Bloch sphere. The photon number scales as the square of the rotation angle as expected since the Rabi frequency scales as the drive amplitude.
The observed difference between $\expval{n_\mathrm{out}}$ for both qubit measurement outcomes is negligible compared to the total number of photons in the pulse, as expected from the strong overlap of states $|\psi_g\rangle$ and $|\psi_e\rangle$.

To reveal the difference between the energies of these states, we thus subtract the mean number of photons contained in the incoming pulse $n_\mathrm{in} = \int_0^{t_\mathrm{d}} |\alpha_{\mathrm{in}}(t)|^2 \dd t$ (Fig. \ref{fig:figure_3}b). Without post-selection, the difference $\Delta n=\expval{n_\mathrm{out}}-n_\mathrm{in}$ oscillates between $-1$ and $0$, as expected from the principle of energy conservation:
when the qubit is excited, it extracts a photon from the pulse and when it is in the ground state the pulse energy stays unchanged. Note that this average loss of one photon owing to energy conservation is not necessarily enforced by the application of the annihilation operator, which could even lead to an increase of the photon number for well chosen quantum states~\cite{Mizrahi2002}.
As commonly observed with weak value measurements, the oscillation amplitudes of the post-selected $\Delta n_{g,e}$ can exceed the non post-selected amplitude (blue and red dots compared to shaded area in Fig. \ref{fig:figure_3}b)~\cite{dressel_colloquium_2014}. The post-selected photon number $\Delta n_g$ oscillates in counter-phase with $\Delta n_e$: the information acquired on the qubit state distorts the probability of finding a given photon number in the drive pulse. 

\begin{figure}
\centering
\includegraphics{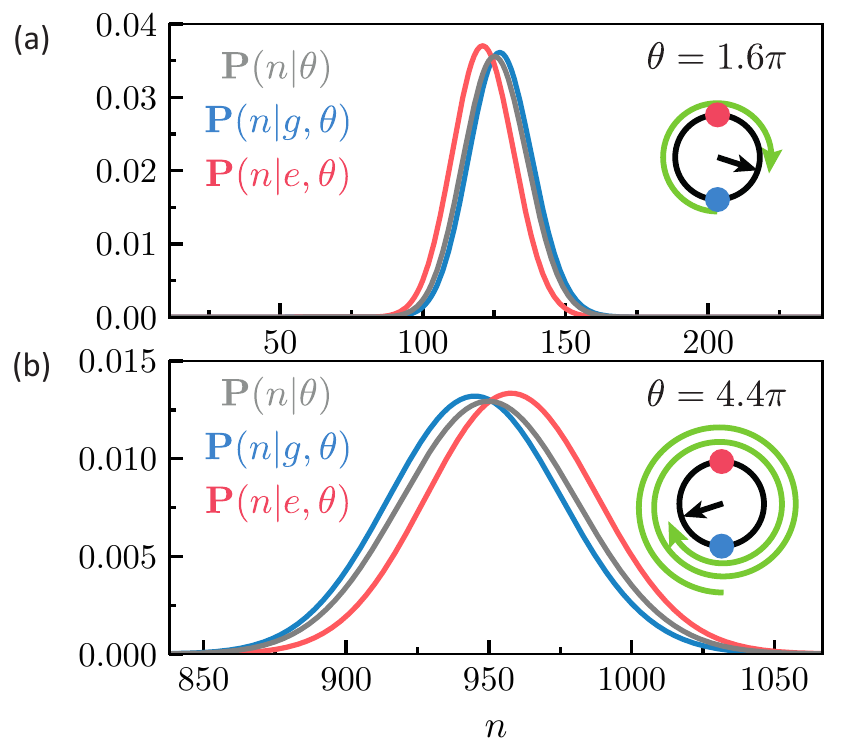}
\caption{Probability distribution that the drive pulse contains $n$ photons knowing that it was prepared in a coherent state leading to a Rabi rotation of $\theta=1.6\pi$ (a) or $\theta=4.4\pi$ (b). Colors encode the post-selected outcome of the qubit measurement: no post-selection (grey), $|g\rangle$ (blue) and $|e\rangle$ (red). 
Insets: Bloch representation of the qubit state after the drive pulse has left it. Green arrow: Rabi rotation. Blue and red dots: $|g\rangle$ and $|e\rangle$ states.
}
\label{fig:figure_4}
\end{figure}
To better understand the effect of the qubit measurement on the photon distribution, we consider a toy model where the drive pulse is modeled as a stationary harmonic oscillator, which interacts with a decoherence free qubit for a time $t_\mathrm{d}$ at a  fixed rate $\Gamma_a$~\cite{Ghosh:97}. A complete description would treat the drive pulse as a propagating field \cite{Silberfarb2003,fischer_scattering_2018,maffei_probing_2021} and yields identical results. The oscillator starts in a coherent state $|\sqrt{n_\mathrm{in}}\rangle=|\theta/\sqrt{4\Gamma_at_\mathrm{d}}\rangle$ with a Poisson distribution $\mathbf{P}_{\theta}(n)$ for the photon number centered on $n_\mathrm{in}$ (grey lines in Fig.~\ref{fig:figure_4}). Post-selecting on a particular qubit measurement outcome distorts this probability distribution. The measurement operators $\hat{M}_g$ and $\hat{M}_e$ describing the backaction exerted on the oscillator when the qubit is measured in $|g\rangle$  or in $|e\rangle$ read $\hat{M}_g = \cos(\sqrt{4\Gamma_a t_\mathrm{d} \hat{a}^\dagger\hat{a}})$ and $\hat{M}_e = \hat{e} \sin(\sqrt{4\Gamma_a t_\mathrm{d} \hat{a}^\dagger\hat{a}})$, where $\hat{e} = \sum_n \ket{n}\bra{n+1}$ is the bare lowering operator (see~\cite{noauthor_supplementary_nodate}). Inspired by the problem of photodetection of a cavity output~\cite{noauthor_supplementary_nodate}, we distinguish two  effects in the backaction: (i) the Bayesian update on the photon distribution conditioned on the measurement outcome and (ii) the extraction of a single photon from the drive pulse which is used to flip the qubit into its excited state.

Through (i), the Poisson distribution are multiplied by $\langle n|\hat{M}_i^\dagger \hat{M}_i|n\rangle$, which is either $\cos^2(\sqrt{n\Gamma_a t_\mathrm{d}})$ or $\sin^2(\sqrt{n\Gamma_a t_\mathrm{d}})$, and then renormalized (see section 9 in~\cite{noauthor_supplementary_nodate}). This Bayesian update leads to an increase or a decrease of the mean occupancy~\cite{Ueda1992,Nunn2022}. 
The direction depends on the rotation angle since the outcome of the qubit measurement indicates that the qubit is either ahead of its average evolution (more photons than expected in the drive), or behind (less photons). One can see that for $\theta=1.6\pi$, finding the qubit in $|g\rangle$ projects it ahead of its average evolution and thus offsets the probability distribution $\mathbf{P}_\theta(n|g)$ towards larger photon numbers. Each half turn, the situation reverses, explaining why for $\theta=4.4\pi$, $\mathbf{P}_\theta(n|g)$ is offset towards smaller photon numbers. This behavior explains the oscillations we observe in Fig.~\ref{fig:figure_3}. Moreover, owing to the increasing standard deviation of the Poisson distribution  $\mathbf{P}_\theta(n)$ with the amplitude $\sqrt{n_\mathrm{in}}\propto\theta$, the backaction on $\Delta n$ increases linearly with $\theta$ (dotted lines in Fig.~\ref{fig:figure_3}b and \cite{noauthor_supplementary_nodate}). 

Through (ii), the qubit measurement backaction entails the destruction of a photon in the drive pulse when the qubit is found in $|e\rangle$ and no extra cost when in $|g\rangle$. This single photon offset corresponds to the operator $\hat{e}$ in $\hat{M}_e$ and amounts to the minimum of the measured oscillations in the non post-selected average $\Delta n$. For the post-selected cases, this contribution of the measurement backaction  is not immediately visible in the measured $\Delta n_{g,e}$, but can be made explicit in the predicted oscillations derived from the past quantum state model of Fig.~\ref{fig:figure_3}b (see~\cite{noauthor_supplementary_nodate}).

In conclusion, we measured the energy flows between a qubit and the resonant drive commonly used to perform single-qubit gates. The unavoidable entanglement between the qubit and the drive reflects on an observable energy exchange. In this context, the projective measurement of the qubit can be understood as a weak measurement of the drive pulse. 
The experiment is therefore able to clearly demonstrate a correlation between the propagating driving pulse and the qubit, which eventually sets an upper bound on the gate fidelity. Ultimately, the kind of measurements we performed illustrate the limitations set by energy conservation on gate fidelity~\cite{Gea_Banacloche_2005}. The energy change of the drive pulse resulting from the qubit measurement can even exceed the maximal qubit extracted energy of one photon. While surprising when considering the average experiment, it is well explained by a weak-value model. Looking forward, it would be interesting to perform a full quantum tomography of the drive state using newly-developed itinerant mode detectors~\cite{besse_parity_2020,dassonneville_number-resolved_2020} by first displacing the quantum state towards low photon numbers, similarly to the steady state case explored in Ref.~\cite{Lu2021}. We indeed expect the driving mode to be in a controllable non-Gaussian state. Using a squeezed drive would also enable to quantify the amount of entanglement between qubit and drive that is qubit state dependent~\cite{Shahmoon2009} and even suppress it fully~\cite{Goldberg2020}. From a thermodynamic point of view, this measurement backaction on the energy is at the core of the class of quantum thermodynamic engines that are powered by measurements instead of heat bath~\cite{brandner_coherence-enhanced_2015,yi_single-temperature_2017,elouard_extracting_2017,elouard_efficient_2018,ding_measurement-driven_2018,abdelkhalek_fundamental_2018,buffoni_quantum_2019,solfanelli_maximal_2019,seah_maxwells_2020,bresque_two-qubit_2021,manikandan_efficiently_2021}. 
Finally, we note that our work can be recast in the framework of quantum batteries~\cite{Campaioli2017,Ferraro2018,Binder2015,Andolina2019,Julia-Farre2020,Caravelli2021,Tirone2021}. From that perspective, we realized the anatomy of a charging event for a single qubit battery.

\begin{acknowledgments} 
This research was supported by grant number FQXi-IAF19-05 from the Foundational Questions Institute Fund, a donor advised fund of Silicon Valley Community Foundation, the Fondation Del Duca, the Templeton World Charity Foundation, Inc (Grant No. TWCF0338), the ANR Research Collaborative Project "Qu-DICE" (ANR-PRC-CES47) and the John Templeton Foundation grant no. 61835. We acknowledge IARPA and Lincoln Labs for providing a Josephson Traveling-Wave Parametric Amplifier. We thank Igor Dotsenko for his useful feedback.
\end{acknowledgments}


\begin{thebibliography}{88}%
\makeatletter
\providecommand \@ifxundefined [1]{%
 \@ifx{#1\undefined}
}%
\providecommand \@ifnum [1]{%
 \ifnum #1\expandafter \@firstoftwo
 \else \expandafter \@secondoftwo
 \fi
}%
\providecommand \@ifx [1]{%
 \ifx #1\expandafter \@firstoftwo
 \else \expandafter \@secondoftwo
 \fi
}%
\providecommand \natexlab [1]{#1}%
\providecommand \enquote  [1]{``#1''}%
\providecommand \bibnamefont  [1]{#1}%
\providecommand \bibfnamefont [1]{#1}%
\providecommand \citenamefont [1]{#1}%
\providecommand \href@noop [0]{\@secondoftwo}%
\providecommand \href [0]{\begingroup \@sanitize@url \@href}%
\providecommand \@href[1]{\@@startlink{#1}\@@href}%
\providecommand \@@href[1]{\endgroup#1\@@endlink}%
\providecommand \@sanitize@url [0]{\catcode `\\12\catcode `\$12\catcode
  `\&12\catcode `\#12\catcode `\^12\catcode `\_12\catcode `\%12\relax}%
\providecommand \@@startlink[1]{}%
\providecommand \@@endlink[0]{}%
\providecommand \url  [0]{\begingroup\@sanitize@url \@url }%
\providecommand \@url [1]{\endgroup\@href {#1}{\urlprefix }}%
\providecommand \urlprefix  [0]{URL }%
\providecommand \Eprint [0]{\href }%
\providecommand \doibase [0]{https://doi.org/}%
\providecommand \selectlanguage [0]{\@gobble}%
\providecommand \bibinfo  [0]{\@secondoftwo}%
\providecommand \bibfield  [0]{\@secondoftwo}%
\providecommand \translation [1]{[#1]}%
\providecommand \BibitemOpen [0]{}%
\providecommand \bibitemStop [0]{}%
\providecommand \bibitemNoStop [0]{.\EOS\space}%
\providecommand \EOS [0]{\spacefactor3000\relax}%
\providecommand \BibitemShut  [1]{\csname bibitem#1\endcsname}%
\let\auto@bib@innerbib\@empty
\bibitem [{\citenamefont {Barnes}\ and\ \citenamefont
  {Warren}(1999)}]{Barnes1999}%
  \BibitemOpen
  \bibfield  {author} {\bibinfo {author} {\bibfnamefont {J.~P.}\ \bibnamefont
  {Barnes}}\ and\ \bibinfo {author} {\bibfnamefont {W.~S.}\ \bibnamefont
  {Warren}},\ }\bibfield  {title} {\bibinfo {title} {{Decoherence and
  programmable quantum computation}},\ }\href
  {https://doi.org/10.1103/PhysRevA.60.4363} {\bibfield  {journal} {\bibinfo
  {journal} {Physical Review A}\ }\textbf {\bibinfo {volume} {60}},\ \bibinfo
  {pages} {4363} (\bibinfo {year} {1999})}\BibitemShut {NoStop}%
\bibitem [{\citenamefont
  {Gea-Banacloche}(2002{\natexlab{a}})}]{gea-banacloche_implications_2002}%
  \BibitemOpen
  \bibfield  {author} {\bibinfo {author} {\bibfnamefont {J.}~\bibnamefont
  {Gea-Banacloche}},\ }\bibfield  {title} {\bibinfo {title} {Some implications
  of the quantum nature of laser fields for quantum computations},\ }\href
  {https://doi.org/10.1103/PhysRevA.65.022308} {\bibfield  {journal} {\bibinfo
  {journal} {Physical Review A}\ }\textbf {\bibinfo {volume} {65}},\ \bibinfo
  {pages} {022308} (\bibinfo {year} {2002}{\natexlab{a}})}\BibitemShut
  {NoStop}%
\bibitem [{\citenamefont
  {Gea-Banacloche}(2002{\natexlab{b}})}]{gea-banacloche_minimum_2002}%
  \BibitemOpen
  \bibfield  {author} {\bibinfo {author} {\bibfnamefont {J.}~\bibnamefont
  {Gea-Banacloche}},\ }\bibfield  {title} {\bibinfo {title} {Minimum {Energy}
  {Requirements} for {Quantum} {Computation}},\ }\href
  {https://doi.org/10.1103/PhysRevLett.89.217901} {\bibfield  {journal}
  {\bibinfo  {journal} {Physical Review Letters}\ }\textbf {\bibinfo {volume}
  {89}},\ \bibinfo {pages} {217901} (\bibinfo {year}
  {2002}{\natexlab{b}})}\BibitemShut {NoStop}%
\bibitem [{\citenamefont {Ozawa}(2002)}]{ozawa_conservative_2002}%
  \BibitemOpen
  \bibfield  {author} {\bibinfo {author} {\bibfnamefont {M.}~\bibnamefont
  {Ozawa}},\ }\bibfield  {title} {\bibinfo {title} {Conservative {Quantum}
  {Computing}},\ }\href {https://doi.org/10.1103/PhysRevLett.89.057902}
  {\bibfield  {journal} {\bibinfo  {journal} {Physical Review Letters}\
  }\textbf {\bibinfo {volume} {89}},\ \bibinfo {pages} {057902} (\bibinfo
  {year} {2002})}\BibitemShut {NoStop}%
\bibitem [{\citenamefont {Karasawa}\ and\ \citenamefont
  {Ozawa}(2007)}]{Karasawa2007}%
  \BibitemOpen
  \bibfield  {author} {\bibinfo {author} {\bibfnamefont {T.}~\bibnamefont
  {Karasawa}}\ and\ \bibinfo {author} {\bibfnamefont {M.}~\bibnamefont
  {Ozawa}},\ }\bibfield  {title} {\bibinfo {title} {{Conservation-law-induced
  quantum limits for physical realizations of the quantum NOT gate}},\ }\href
  {https://doi.org/10.1103/PhysRevA.75.032324} {\bibfield  {journal} {\bibinfo
  {journal} {Physical Review A}\ }\textbf {\bibinfo {volume} {75}},\ \bibinfo
  {pages} {032324} (\bibinfo {year} {2007})}\BibitemShut {NoStop}%
\bibitem [{\citenamefont {Bedingham}\ and\ \citenamefont
  {Maroney}(2016)}]{Bedingham16}%
  \BibitemOpen
  \bibfield  {author} {\bibinfo {author} {\bibfnamefont {D.~J.}\ \bibnamefont
  {Bedingham}}\ and\ \bibinfo {author} {\bibfnamefont {O.~J.~E.}\ \bibnamefont
  {Maroney}},\ }\bibfield  {title} {\bibinfo {title} {{The thermodynamic cost
  of quantum operations}},\ }\href
  {https://doi.org/10.1088/1367-2630/18/11/113050} {\bibfield  {journal}
  {\bibinfo  {journal} {New J. Phys.}\ }\textbf {\bibinfo {volume} {18}},\
  \bibinfo {pages} {113050} (\bibinfo {year} {2016})}\BibitemShut {NoStop}%
\bibitem [{\citenamefont {Ikonen}\ \emph {et~al.}(2017)\citenamefont {Ikonen},
  \citenamefont {Salmilehto},\ and\ \citenamefont
  {Möttönen}}]{ikonen_energy-efficient_2017}%
  \BibitemOpen
  \bibfield  {author} {\bibinfo {author} {\bibfnamefont {J.}~\bibnamefont
  {Ikonen}}, \bibinfo {author} {\bibfnamefont {J.}~\bibnamefont {Salmilehto}},\
  and\ \bibinfo {author} {\bibfnamefont {M.}~\bibnamefont {Möttönen}},\
  }\bibfield  {title} {\bibinfo {title} {Energy-efficient quantum computing},\
  }\href {https://doi.org/10.1038/s41534-017-0015-5} {\bibfield  {journal}
  {\bibinfo  {journal} {npj Quantum Information}\ }\textbf {\bibinfo {volume}
  {3}},\ \bibinfo {pages} {1} (\bibinfo {year} {2017})}\BibitemShut {NoStop}%
\bibitem [{\citenamefont {Guryanova}\ \emph {et~al.}(2020)\citenamefont
  {Guryanova}, \citenamefont {Friis},\ and\ \citenamefont
  {Huber}}]{Guryanova20}%
  \BibitemOpen
  \bibfield  {author} {\bibinfo {author} {\bibfnamefont {Y.}~\bibnamefont
  {Guryanova}}, \bibinfo {author} {\bibfnamefont {N.}~\bibnamefont {Friis}},\
  and\ \bibinfo {author} {\bibfnamefont {M.}~\bibnamefont {Huber}},\ }\bibfield
   {title} {\bibinfo {title} {{Ideal Projective Measurements Have Infinite
  Resource Costs}},\ }\href {https://doi.org/10.22331/q-2020-01-13-222}
  {\bibfield  {journal} {\bibinfo  {journal} {Quantum}\ }\textbf {\bibinfo
  {volume} {4}},\ \bibinfo {pages} {222} (\bibinfo {year} {2020})}\BibitemShut
  {NoStop}%
\bibitem [{\citenamefont {Cimini}\ \emph {et~al.}(2020)\citenamefont {Cimini},
  \citenamefont {Gherardini}, \citenamefont {Barbieri}, \citenamefont
  {Gianani}, \citenamefont {Sbroscia}, \citenamefont {Buffoni}, \citenamefont
  {Paternostro},\ and\ \citenamefont {Caruso}}]{Cimini2020}%
  \BibitemOpen
  \bibfield  {author} {\bibinfo {author} {\bibfnamefont {V.}~\bibnamefont
  {Cimini}}, \bibinfo {author} {\bibfnamefont {S.}~\bibnamefont {Gherardini}},
  \bibinfo {author} {\bibfnamefont {M.}~\bibnamefont {Barbieri}}, \bibinfo
  {author} {\bibfnamefont {I.}~\bibnamefont {Gianani}}, \bibinfo {author}
  {\bibfnamefont {M.}~\bibnamefont {Sbroscia}}, \bibinfo {author}
  {\bibfnamefont {L.}~\bibnamefont {Buffoni}}, \bibinfo {author} {\bibfnamefont
  {M.}~\bibnamefont {Paternostro}},\ and\ \bibinfo {author} {\bibfnamefont
  {F.}~\bibnamefont {Caruso}},\ }\bibfield  {title} {\bibinfo {title}
  {{Experimental characterization of the energetics of quantum logic gates}},\
  }\href {https://doi.org/10.1038/s41534-020-00325-7} {\bibfield  {journal}
  {\bibinfo  {journal} {npj Quantum Information}\ }\textbf {\bibinfo {volume}
  {6}},\ \bibinfo {pages} {96} (\bibinfo {year} {2020})}\BibitemShut {NoStop}%
\bibitem [{\citenamefont {Deffner}(2021)}]{Deffner2021}%
  \BibitemOpen
  \bibfield  {author} {\bibinfo {author} {\bibfnamefont {S.}~\bibnamefont
  {Deffner}},\ }\bibfield  {title} {\bibinfo {title} {{Energetic cost of
  Hamiltonian quantum gates}},\ }\href
  {https://doi.org/10.1209/0295-5075/134/40002} {\bibfield  {journal} {\bibinfo
   {journal} {Europhys. Lett.}\ }\textbf {\bibinfo {volume} {134}},\ \bibinfo
  {pages} {40002} (\bibinfo {year} {2021})}\BibitemShut {NoStop}%
\bibitem [{\citenamefont {Faist}\ \emph {et~al.}(2015)\citenamefont {Faist},
  \citenamefont {Dupuis}, \citenamefont {Oppenheim},\ and\ \citenamefont
  {Renner}}]{Faist15}%
  \BibitemOpen
  \bibfield  {author} {\bibinfo {author} {\bibfnamefont {P.}~\bibnamefont
  {Faist}}, \bibinfo {author} {\bibfnamefont {F.}~\bibnamefont {Dupuis}},
  \bibinfo {author} {\bibfnamefont {J.}~\bibnamefont {Oppenheim}},\ and\
  \bibinfo {author} {\bibfnamefont {R.}~\bibnamefont {Renner}},\ }\bibfield
  {title} {\bibinfo {title} {{The minimal work cost of information processing -
  Nature Communications}},\ }\href {https://doi.org/10.1038/ncomms8669}
  {\bibfield  {journal} {\bibinfo  {journal} {Nat. Commun.}\ }\textbf {\bibinfo
  {volume} {6}},\ \bibinfo {pages} {1} (\bibinfo {year} {2015})}\BibitemShut
  {NoStop}%
\bibitem [{\citenamefont {Reeb}\ and\ \citenamefont {Wolf}(2014)}]{Reeb14}%
  \BibitemOpen
  \bibfield  {author} {\bibinfo {author} {\bibfnamefont {D.}~\bibnamefont
  {Reeb}}\ and\ \bibinfo {author} {\bibfnamefont {M.~M.}\ \bibnamefont
  {Wolf}},\ }\bibfield  {title} {\bibinfo {title} {{An improved Landauer
  principle with finite-size corrections}},\ }\href
  {https://doi.org/10.1088/1367-2630/16/10/103011} {\bibfield  {journal}
  {\bibinfo  {journal} {New J. Phys.}\ }\textbf {\bibinfo {volume} {16}},\
  \bibinfo {pages} {103011} (\bibinfo {year} {2014})}\BibitemShut {NoStop}%
\bibitem [{\citenamefont {Sagawa}\ and\ \citenamefont {Ueda}(2009)}]{Sagawa09}%
  \BibitemOpen
  \bibfield  {author} {\bibinfo {author} {\bibfnamefont {T.}~\bibnamefont
  {Sagawa}}\ and\ \bibinfo {author} {\bibfnamefont {M.}~\bibnamefont {Ueda}},\
  }\bibfield  {title} {\bibinfo {title} {{Minimal Energy Cost for Thermodynamic
  Information Processing: Measurement and Information Erasure}},\ }\href
  {https://doi.org/10.1103/PhysRevLett.102.250602} {\bibfield  {journal}
  {\bibinfo  {journal} {Phys. Rev. Lett.}\ }\textbf {\bibinfo {volume} {102}},\
  \bibinfo {pages} {250602} (\bibinfo {year} {2009})}\BibitemShut {NoStop}%
\bibitem [{\citenamefont {Mohammady}\ and\ \citenamefont
  {Romito}(2019)}]{Mohammady19}%
  \BibitemOpen
  \bibfield  {author} {\bibinfo {author} {\bibfnamefont {M.~H.}\ \bibnamefont
  {Mohammady}}\ and\ \bibinfo {author} {\bibfnamefont {A.}~\bibnamefont
  {Romito}},\ }\bibfield  {title} {\bibinfo {title} {{Conditional work
  statistics of quantum measurements}},\ }\href
  {https://doi.org/10.22331/q-2019-08-19-175} {\bibfield  {journal} {\bibinfo
  {journal} {Quantum}\ }\textbf {\bibinfo {volume} {3}},\ \bibinfo {pages}
  {175} (\bibinfo {year} {2019})}\BibitemShut {NoStop}%
\bibitem [{\citenamefont {Aifer}\ and\ \citenamefont
  {Deffner}(2022)}]{Aifer2022}%
  \BibitemOpen
  \bibfield  {author} {\bibinfo {author} {\bibfnamefont {M.}~\bibnamefont
  {Aifer}}\ and\ \bibinfo {author} {\bibfnamefont {S.}~\bibnamefont
  {Deffner}},\ }\bibfield  {title} {\bibinfo {title} {From quantum speed limits
  to energy-efficient quantum gates},\ }\href
  {https://doi.org/10.1088/1367-2630/ac6821} {\bibfield  {journal} {\bibinfo
  {journal} {New Journal of Physics}\ }\textbf {\bibinfo {volume} {24}},\
  \bibinfo {pages} {055002} (\bibinfo {year} {2022})}\BibitemShut {NoStop}%
\bibitem [{\citenamefont {Gu}\ \emph {et~al.}(2017)\citenamefont {Gu},
  \citenamefont {Kockum}, \citenamefont {Miranowicz}, \citenamefont {Liu},\
  and\ \citenamefont {Nori}}]{gu_microwave_2017}%
  \BibitemOpen
  \bibfield  {author} {\bibinfo {author} {\bibfnamefont {X.}~\bibnamefont
  {Gu}}, \bibinfo {author} {\bibfnamefont {A.~F.}\ \bibnamefont {Kockum}},
  \bibinfo {author} {\bibfnamefont {A.}~\bibnamefont {Miranowicz}}, \bibinfo
  {author} {\bibfnamefont {Y.-x.}\ \bibnamefont {Liu}},\ and\ \bibinfo {author}
  {\bibfnamefont {F.}~\bibnamefont {Nori}},\ }\bibfield  {title} {\bibinfo
  {title} {Microwave photonics with superconducting quantum circuits},\ }\href
  {https://doi.org/10.1016/j.physrep.2017.10.002} {\bibfield  {journal}
  {\bibinfo  {journal} {Physics Reports}\ }\bibinfo {series} {Microwave
  photonics with superconducting quantum circuits},\ \textbf {\bibinfo {volume}
  {718-719}},\ \bibinfo {pages} {1} (\bibinfo {year} {2017})}\BibitemShut
  {NoStop}%
\bibitem [{\citenamefont {Houck}\ \emph {et~al.}(2007)\citenamefont {Houck},
  \citenamefont {Schuster}, \citenamefont {Gambetta}, \citenamefont {Schreier},
  \citenamefont {Johnson}, \citenamefont {Chow}, \citenamefont {Frunzio},
  \citenamefont {Majer}, \citenamefont {Devoret}, \citenamefont {Girvin},\ and\
  \citenamefont {Schoelkopf}}]{houck_generating_2007}%
  \BibitemOpen
  \bibfield  {author} {\bibinfo {author} {\bibfnamefont {A.~A.}\ \bibnamefont
  {Houck}}, \bibinfo {author} {\bibfnamefont {D.~I.}\ \bibnamefont {Schuster}},
  \bibinfo {author} {\bibfnamefont {J.~M.}\ \bibnamefont {Gambetta}}, \bibinfo
  {author} {\bibfnamefont {J.~A.}\ \bibnamefont {Schreier}}, \bibinfo {author}
  {\bibfnamefont {B.~R.}\ \bibnamefont {Johnson}}, \bibinfo {author}
  {\bibfnamefont {J.~M.}\ \bibnamefont {Chow}}, \bibinfo {author}
  {\bibfnamefont {L.}~\bibnamefont {Frunzio}}, \bibinfo {author} {\bibfnamefont
  {J.}~\bibnamefont {Majer}}, \bibinfo {author} {\bibfnamefont {M.~H.}\
  \bibnamefont {Devoret}}, \bibinfo {author} {\bibfnamefont {S.~M.}\
  \bibnamefont {Girvin}},\ and\ \bibinfo {author} {\bibfnamefont {R.~J.}\
  \bibnamefont {Schoelkopf}},\ }\bibfield  {title} {\bibinfo {title}
  {Generating single microwave photons in a circuit},\ }\href
  {https://doi.org/10.1038/nature06126} {\bibfield  {journal} {\bibinfo
  {journal} {Nature}\ }\textbf {\bibinfo {volume} {449}},\ \bibinfo {pages}
  {328} (\bibinfo {year} {2007})}\BibitemShut {NoStop}%
\bibitem [{\citenamefont {Hoi}\ \emph {et~al.}(2011)\citenamefont {Hoi},
  \citenamefont {Wilson}, \citenamefont {Johansson}, \citenamefont {Palomaki},
  \citenamefont {Peropadre},\ and\ \citenamefont
  {Delsing}}]{hoi_demonstration_2011}%
  \BibitemOpen
  \bibfield  {author} {\bibinfo {author} {\bibfnamefont {I.-C.}\ \bibnamefont
  {Hoi}}, \bibinfo {author} {\bibfnamefont {C.~M.}\ \bibnamefont {Wilson}},
  \bibinfo {author} {\bibfnamefont {G.}~\bibnamefont {Johansson}}, \bibinfo
  {author} {\bibfnamefont {T.}~\bibnamefont {Palomaki}}, \bibinfo {author}
  {\bibfnamefont {B.}~\bibnamefont {Peropadre}},\ and\ \bibinfo {author}
  {\bibfnamefont {P.}~\bibnamefont {Delsing}},\ }\bibfield  {title} {\bibinfo
  {title} {Demonstration of a {Single}-{Photon} {Router} in the {Microwave}
  {Regime}},\ }\href {https://doi.org/10.1103/PhysRevLett.107.073601}
  {\bibfield  {journal} {\bibinfo  {journal} {Physical Review Letters}\
  }\textbf {\bibinfo {volume} {107}},\ \bibinfo {pages} {073601} (\bibinfo
  {year} {2011})}\BibitemShut {NoStop}%
\bibitem [{\citenamefont {Hoi}\ \emph {et~al.}(2012)\citenamefont {Hoi},
  \citenamefont {Palomaki}, \citenamefont {Lindkvist}, \citenamefont
  {Johansson}, \citenamefont {Delsing},\ and\ \citenamefont
  {Wilson}}]{hoi_generation_2012}%
  \BibitemOpen
  \bibfield  {author} {\bibinfo {author} {\bibfnamefont {I.-C.}\ \bibnamefont
  {Hoi}}, \bibinfo {author} {\bibfnamefont {T.}~\bibnamefont {Palomaki}},
  \bibinfo {author} {\bibfnamefont {J.}~\bibnamefont {Lindkvist}}, \bibinfo
  {author} {\bibfnamefont {G.}~\bibnamefont {Johansson}}, \bibinfo {author}
  {\bibfnamefont {P.}~\bibnamefont {Delsing}},\ and\ \bibinfo {author}
  {\bibfnamefont {C.~M.}\ \bibnamefont {Wilson}},\ }\bibfield  {title}
  {\bibinfo {title} {Generation of {Nonclassical} {Microwave} {States} {Using}
  an {Artificial} {Atom} in {1D} {Open} {Space}},\ }\href
  {https://doi.org/10.1103/PhysRevLett.108.263601} {\bibfield  {journal}
  {\bibinfo  {journal} {Physical Review Letters}\ }\textbf {\bibinfo {volume}
  {108}},\ \bibinfo {pages} {263601} (\bibinfo {year} {2012})}\BibitemShut
  {NoStop}%
\bibitem [{\citenamefont {Reuer}\ \emph {et~al.}(2021)\citenamefont {Reuer},
  \citenamefont {Besse}, \citenamefont {Wernli}, \citenamefont {Magnard},
  \citenamefont {Kurpiers}, \citenamefont {Norris}, \citenamefont {Wallraff},\
  and\ \citenamefont {Eichler}}]{reuer_realization_2021}%
  \BibitemOpen
  \bibfield  {author} {\bibinfo {author} {\bibfnamefont {K.}~\bibnamefont
  {Reuer}}, \bibinfo {author} {\bibfnamefont {J.-C.}\ \bibnamefont {Besse}},
  \bibinfo {author} {\bibfnamefont {L.}~\bibnamefont {Wernli}}, \bibinfo
  {author} {\bibfnamefont {P.}~\bibnamefont {Magnard}}, \bibinfo {author}
  {\bibfnamefont {P.}~\bibnamefont {Kurpiers}}, \bibinfo {author}
  {\bibfnamefont {G.~J.}\ \bibnamefont {Norris}}, \bibinfo {author}
  {\bibfnamefont {A.}~\bibnamefont {Wallraff}},\ and\ \bibinfo {author}
  {\bibfnamefont {C.}~\bibnamefont {Eichler}},\ }\bibfield  {title} {\bibinfo
  {title} {Realization of a {Universal} {Quantum} {Gate} {Set} for {Itinerant}
  {Microwave} {Photons}},\ }\href {http://arxiv.org/abs/2106.03481} {\bibfield
  {journal} {\bibinfo  {journal} {arXiv:2106.03481 [quant-ph]}\ } (\bibinfo
  {year} {2021})}\BibitemShut {NoStop}%
\bibitem [{\citenamefont {Astafiev}\ \emph {et~al.}(2010)\citenamefont
  {Astafiev}, \citenamefont {Zagoskin}, \citenamefont {Abdumalikov},
  \citenamefont {Pashkin}, \citenamefont {Yamamoto}, \citenamefont {Inomata},
  \citenamefont {Nakamura},\ and\ \citenamefont
  {Tsai}}]{astafiev_resonance_2010}%
  \BibitemOpen
  \bibfield  {author} {\bibinfo {author} {\bibfnamefont {O.}~\bibnamefont
  {Astafiev}}, \bibinfo {author} {\bibfnamefont {A.~M.}\ \bibnamefont
  {Zagoskin}}, \bibinfo {author} {\bibfnamefont {A.~A.}\ \bibnamefont
  {Abdumalikov}}, \bibinfo {author} {\bibfnamefont {Y.~A.}\ \bibnamefont
  {Pashkin}}, \bibinfo {author} {\bibfnamefont {T.}~\bibnamefont {Yamamoto}},
  \bibinfo {author} {\bibfnamefont {K.}~\bibnamefont {Inomata}}, \bibinfo
  {author} {\bibfnamefont {Y.}~\bibnamefont {Nakamura}},\ and\ \bibinfo
  {author} {\bibfnamefont {J.~S.}\ \bibnamefont {Tsai}},\ }\bibfield  {title}
  {\bibinfo {title} {Resonance {Fluorescence} of a {Single} {Artificial}
  {Atom}},\ }\href {https://doi.org/10.1126/science.1181918} {\bibfield
  {journal} {\bibinfo  {journal} {Science}\ }\textbf {\bibinfo {volume}
  {327}},\ \bibinfo {pages} {840} (\bibinfo {year} {2010})}\BibitemShut
  {NoStop}%
\bibitem [{\citenamefont {Abdumalikov}\ \emph {et~al.}(2011)\citenamefont
  {Abdumalikov}, \citenamefont {Astafiev}, \citenamefont {Pashkin},
  \citenamefont {Nakamura},\ and\ \citenamefont
  {Tsai}}]{abdumalikov_dynamics_2011}%
  \BibitemOpen
  \bibfield  {author} {\bibinfo {author} {\bibfnamefont {A.~A.}\ \bibnamefont
  {Abdumalikov}}, \bibinfo {author} {\bibfnamefont {O.~V.}\ \bibnamefont
  {Astafiev}}, \bibinfo {author} {\bibfnamefont {Y.~A.}\ \bibnamefont
  {Pashkin}}, \bibinfo {author} {\bibfnamefont {Y.}~\bibnamefont {Nakamura}},\
  and\ \bibinfo {author} {\bibfnamefont {J.~S.}\ \bibnamefont {Tsai}},\
  }\bibfield  {title} {\bibinfo {title} {Dynamics of {Coherent} and
  {Incoherent} {Emission} from an {Artificial} {Atom} in a {1D} {Space}},\
  }\href {https://doi.org/10.1103/PhysRevLett.107.043604} {\bibfield  {journal}
  {\bibinfo  {journal} {Physical Review Letters}\ }\textbf {\bibinfo {volume}
  {107}},\ \bibinfo {pages} {043604} (\bibinfo {year} {2011})}\BibitemShut
  {NoStop}%
\bibitem [{\citenamefont {Hönigl-Decrinis}\ \emph {et~al.}(2020)\citenamefont
  {Hönigl-Decrinis}, \citenamefont {Shaikhaidarov}, \citenamefont {de~Graaf},
  \citenamefont {Antonov},\ and\ \citenamefont
  {Astafiev}}]{honigl-decrinis_two-level_2020}%
  \BibitemOpen
  \bibfield  {author} {\bibinfo {author} {\bibfnamefont {T.}~\bibnamefont
  {Hönigl-Decrinis}}, \bibinfo {author} {\bibfnamefont {R.}~\bibnamefont
  {Shaikhaidarov}}, \bibinfo {author} {\bibfnamefont {S.}~\bibnamefont
  {de~Graaf}}, \bibinfo {author} {\bibfnamefont {V.}~\bibnamefont {Antonov}},\
  and\ \bibinfo {author} {\bibfnamefont {O.}~\bibnamefont {Astafiev}},\
  }\bibfield  {title} {\bibinfo {title} {Two-{Level} {System} as a {Quantum}
  {Sensor} for {Absolute} {Calibration} of {Power}},\ }\href
  {https://doi.org/10.1103/PhysRevApplied.13.024066} {\bibfield  {journal}
  {\bibinfo  {journal} {Physical Review Applied}\ }\textbf {\bibinfo {volume}
  {13}},\ \bibinfo {pages} {024066} (\bibinfo {year} {2020})}\BibitemShut
  {NoStop}%
\bibitem [{\citenamefont {Lu}\ \emph {et~al.}(2021{\natexlab{a}})\citenamefont
  {Lu}, \citenamefont {Bengtsson}, \citenamefont {Burnett}, \citenamefont
  {Wiegand}, \citenamefont {Suri}, \citenamefont {Krantz}, \citenamefont
  {Roudsari}, \citenamefont {Kockum}, \citenamefont {Gasparinetti},
  \citenamefont {Johansson},\ and\ \citenamefont
  {Delsing}}]{lu_characterizing_2021}%
  \BibitemOpen
  \bibfield  {author} {\bibinfo {author} {\bibfnamefont {Y.}~\bibnamefont
  {Lu}}, \bibinfo {author} {\bibfnamefont {A.}~\bibnamefont {Bengtsson}},
  \bibinfo {author} {\bibfnamefont {J.~J.}\ \bibnamefont {Burnett}}, \bibinfo
  {author} {\bibfnamefont {E.}~\bibnamefont {Wiegand}}, \bibinfo {author}
  {\bibfnamefont {B.}~\bibnamefont {Suri}}, \bibinfo {author} {\bibfnamefont
  {P.}~\bibnamefont {Krantz}}, \bibinfo {author} {\bibfnamefont {A.~F.}\
  \bibnamefont {Roudsari}}, \bibinfo {author} {\bibfnamefont {A.~F.}\
  \bibnamefont {Kockum}}, \bibinfo {author} {\bibfnamefont {S.}~\bibnamefont
  {Gasparinetti}}, \bibinfo {author} {\bibfnamefont {G.}~\bibnamefont
  {Johansson}},\ and\ \bibinfo {author} {\bibfnamefont {P.}~\bibnamefont
  {Delsing}},\ }\bibfield  {title} {\bibinfo {title} {Characterizing
  decoherence rates of a superconducting qubit by direct microwave
  scattering},\ }\href {https://doi.org/10.1038/s41534-021-00367-5} {\bibfield
  {journal} {\bibinfo  {journal} {npj Quantum Information}\ }\textbf {\bibinfo
  {volume} {7}},\ \bibinfo {pages} {1} (\bibinfo {year}
  {2021}{\natexlab{a}})}\BibitemShut {NoStop}%
\bibitem [{\citenamefont {Cottet}\ \emph {et~al.}(2017)\citenamefont {Cottet},
  \citenamefont {Jezouin}, \citenamefont {Bretheau}, \citenamefont
  {Campagne-Ibarcq}, \citenamefont {Ficheux}, \citenamefont {Anders},
  \citenamefont {Auffèves}, \citenamefont {Azouit}, \citenamefont {Rouchon},\
  and\ \citenamefont {Huard}}]{cottet_observing_2017}%
  \BibitemOpen
  \bibfield  {author} {\bibinfo {author} {\bibfnamefont {N.}~\bibnamefont
  {Cottet}}, \bibinfo {author} {\bibfnamefont {S.}~\bibnamefont {Jezouin}},
  \bibinfo {author} {\bibfnamefont {L.}~\bibnamefont {Bretheau}}, \bibinfo
  {author} {\bibfnamefont {P.}~\bibnamefont {Campagne-Ibarcq}}, \bibinfo
  {author} {\bibfnamefont {Q.}~\bibnamefont {Ficheux}}, \bibinfo {author}
  {\bibfnamefont {J.}~\bibnamefont {Anders}}, \bibinfo {author} {\bibfnamefont
  {A.}~\bibnamefont {Auffèves}}, \bibinfo {author} {\bibfnamefont
  {R.}~\bibnamefont {Azouit}}, \bibinfo {author} {\bibfnamefont
  {P.}~\bibnamefont {Rouchon}},\ and\ \bibinfo {author} {\bibfnamefont
  {B.}~\bibnamefont {Huard}},\ }\bibfield  {title} {\bibinfo {title} {Observing
  a quantum {Maxwell} demon at work},\ }\href
  {https://doi.org/10.1073/pnas.1704827114} {\bibfield  {journal} {\bibinfo
  {journal} {Proceedings of the National Academy of Sciences}\ }\textbf
  {\bibinfo {volume} {114}},\ \bibinfo {pages} {7561} (\bibinfo {year}
  {2017})}\BibitemShut {NoStop}%
\bibitem [{\citenamefont {Naghiloo}\ \emph {et~al.}(2020)\citenamefont
  {Naghiloo}, \citenamefont {Tan}, \citenamefont {Harrington}, \citenamefont
  {Alonso}, \citenamefont {Lutz}, \citenamefont {Romito},\ and\ \citenamefont
  {Murch}}]{naghiloo_heat_2020}%
  \BibitemOpen
  \bibfield  {author} {\bibinfo {author} {\bibfnamefont {M.}~\bibnamefont
  {Naghiloo}}, \bibinfo {author} {\bibfnamefont {D.}~\bibnamefont {Tan}},
  \bibinfo {author} {\bibfnamefont {P.}~\bibnamefont {Harrington}}, \bibinfo
  {author} {\bibfnamefont {J.}~\bibnamefont {Alonso}}, \bibinfo {author}
  {\bibfnamefont {E.}~\bibnamefont {Lutz}}, \bibinfo {author} {\bibfnamefont
  {A.}~\bibnamefont {Romito}},\ and\ \bibinfo {author} {\bibfnamefont
  {K.}~\bibnamefont {Murch}},\ }\bibfield  {title} {\bibinfo {title} {Heat and
  {Work} {Along} {Individual} {Trajectories} of a {Quantum} {Bit}},\ }\href
  {https://doi.org/10.1103/PhysRevLett.124.110604} {\bibfield  {journal}
  {\bibinfo  {journal} {Physical Review Letters}\ }\textbf {\bibinfo {volume}
  {124}},\ \bibinfo {pages} {110604} (\bibinfo {year} {2020})}\BibitemShut
  {NoStop}%
\bibitem [{\citenamefont {Binder}\ \emph {et~al.}(2018)\citenamefont {Binder},
  \citenamefont {Correa}, \citenamefont {Gogolin}, \citenamefont {Anders},\
  and\ \citenamefont {Adesso}}]{binder_thermodynamics_2018}%
  \BibitemOpen
  \bibfield  {author} {\bibinfo {author} {\bibfnamefont {F.}~\bibnamefont
  {Binder}}, \bibinfo {author} {\bibfnamefont {L.~A.}\ \bibnamefont {Correa}},
  \bibinfo {author} {\bibfnamefont {C.}~\bibnamefont {Gogolin}}, \bibinfo
  {author} {\bibfnamefont {J.}~\bibnamefont {Anders}},\ and\ \bibinfo {author}
  {\bibfnamefont {G.}~\bibnamefont {Adesso}},\ }\href@noop {} {\emph {\bibinfo
  {title} {Thermodynamics in the {Quantum} {Regime}}}},\ Fundamental {Theories}
  of {Physics}\ (\bibinfo  {publisher} {Springer},\ \bibinfo {year}
  {2018})\BibitemShut {NoStop}%
\bibitem [{\citenamefont {Lu}\ \emph {et~al.}(2021{\natexlab{b}})\citenamefont
  {Lu}, \citenamefont {Strandberg}, \citenamefont {Quijandr{\'{i}}a},
  \citenamefont {Johansson}, \citenamefont {Gasparinetti},\ and\ \citenamefont
  {Delsing}}]{Lu2021}%
  \BibitemOpen
  \bibfield  {author} {\bibinfo {author} {\bibfnamefont {Y.}~\bibnamefont
  {Lu}}, \bibinfo {author} {\bibfnamefont {I.}~\bibnamefont {Strandberg}},
  \bibinfo {author} {\bibfnamefont {F.}~\bibnamefont {Quijandr{\'{i}}a}},
  \bibinfo {author} {\bibfnamefont {G.}~\bibnamefont {Johansson}}, \bibinfo
  {author} {\bibfnamefont {S.}~\bibnamefont {Gasparinetti}},\ and\ \bibinfo
  {author} {\bibfnamefont {P.}~\bibnamefont {Delsing}},\ }\bibfield  {title}
  {\bibinfo {title} {{Propagating Wigner-Negative States Generated from the
  Steady-State Emission of a Superconducting Qubit}},\ }\href
  {https://doi.org/10.1103/PhysRevLett.126.253602} {\bibfield  {journal}
  {\bibinfo  {journal} {Physical Review Letters}\ }\textbf {\bibinfo {volume}
  {126}},\ \bibinfo {pages} {253602} (\bibinfo {year}
  {2021}{\natexlab{b}})}\BibitemShut {NoStop}%
\bibitem [{\citenamefont {Ronzani}\ \emph {et~al.}(2018)\citenamefont
  {Ronzani}, \citenamefont {Karimi}, \citenamefont {Senior}, \citenamefont
  {Chang}, \citenamefont {Peltonen}, \citenamefont {Chen},\ and\ \citenamefont
  {Pekola}}]{ronzani_tunable_2018}%
  \BibitemOpen
  \bibfield  {author} {\bibinfo {author} {\bibfnamefont {A.}~\bibnamefont
  {Ronzani}}, \bibinfo {author} {\bibfnamefont {B.}~\bibnamefont {Karimi}},
  \bibinfo {author} {\bibfnamefont {J.}~\bibnamefont {Senior}}, \bibinfo
  {author} {\bibfnamefont {Y.-C.}\ \bibnamefont {Chang}}, \bibinfo {author}
  {\bibfnamefont {J.~T.}\ \bibnamefont {Peltonen}}, \bibinfo {author}
  {\bibfnamefont {C.}~\bibnamefont {Chen}},\ and\ \bibinfo {author}
  {\bibfnamefont {J.~P.}\ \bibnamefont {Pekola}},\ }\bibfield  {title}
  {\bibinfo {title} {Tunable photonic heat transport in a quantum heat valve},\
  }\href {https://doi.org/10.1038/s41567-018-0199-4} {\bibfield  {journal}
  {\bibinfo  {journal} {Nature Physics}\ }\textbf {\bibinfo {volume} {14}},\
  \bibinfo {pages} {991} (\bibinfo {year} {2018})}\BibitemShut {NoStop}%
\bibitem [{\citenamefont {Senior}\ \emph {et~al.}(2020)\citenamefont {Senior},
  \citenamefont {Gubaydullin}, \citenamefont {Karimi}, \citenamefont
  {Peltonen}, \citenamefont {Ankerhold},\ and\ \citenamefont
  {Pekola}}]{senior_heat_2020}%
  \BibitemOpen
  \bibfield  {author} {\bibinfo {author} {\bibfnamefont {J.}~\bibnamefont
  {Senior}}, \bibinfo {author} {\bibfnamefont {A.}~\bibnamefont {Gubaydullin}},
  \bibinfo {author} {\bibfnamefont {B.}~\bibnamefont {Karimi}}, \bibinfo
  {author} {\bibfnamefont {J.~T.}\ \bibnamefont {Peltonen}}, \bibinfo {author}
  {\bibfnamefont {J.}~\bibnamefont {Ankerhold}},\ and\ \bibinfo {author}
  {\bibfnamefont {J.~P.}\ \bibnamefont {Pekola}},\ }\bibfield  {title}
  {\bibinfo {title} {Heat rectification via a superconducting artificial
  atom},\ }\href {https://doi.org/10.1038/s42005-020-0307-5} {\bibfield
  {journal} {\bibinfo  {journal} {Communications Physics}\ }\textbf {\bibinfo
  {volume} {3}},\ \bibinfo {pages} {1} (\bibinfo {year} {2020})}\BibitemShut
  {NoStop}%
\bibitem [{\citenamefont {Lu}\ \emph {et~al.}(2022)\citenamefont {Lu},
  \citenamefont {Lambert}, \citenamefont {Kockum}, \citenamefont {Funo},
  \citenamefont {Bengtsson}, \citenamefont {Gasparinetti}, \citenamefont
  {Nori},\ and\ \citenamefont {Delsing}}]{Lu2022}%
  \BibitemOpen
  \bibfield  {author} {\bibinfo {author} {\bibfnamefont {Y.}~\bibnamefont
  {Lu}}, \bibinfo {author} {\bibfnamefont {N.}~\bibnamefont {Lambert}},
  \bibinfo {author} {\bibfnamefont {A.~F.}\ \bibnamefont {Kockum}}, \bibinfo
  {author} {\bibfnamefont {K.}~\bibnamefont {Funo}}, \bibinfo {author}
  {\bibfnamefont {A.}~\bibnamefont {Bengtsson}}, \bibinfo {author}
  {\bibfnamefont {S.}~\bibnamefont {Gasparinetti}}, \bibinfo {author}
  {\bibfnamefont {F.}~\bibnamefont {Nori}},\ and\ \bibinfo {author}
  {\bibfnamefont {P.}~\bibnamefont {Delsing}},\ }\bibfield  {title} {\bibinfo
  {title} {{Steady-State Heat Transport and Work With a Single Artificial Atom
  Coupled to a Waveguide: Emission Without External Driving}},\ }\href
  {https://doi.org/10.1103/PRXQuantum.3.020305} {\bibfield  {journal} {\bibinfo
   {journal} {PRX Quantum}\ }\textbf {\bibinfo {volume} {10}},\ \bibinfo
  {pages} {020305} (\bibinfo {year} {2022})}\BibitemShut {NoStop}%
\bibitem [{\citenamefont {Campagne-Ibarcq}\ \emph {et~al.}(2016)\citenamefont
  {Campagne-Ibarcq}, \citenamefont {Six}, \citenamefont {Bretheau},
  \citenamefont {Sarlette}, \citenamefont {Mirrahimi}, \citenamefont
  {Rouchon},\ and\ \citenamefont {Huard}}]{campagne-ibarcq_observing_2016}%
  \BibitemOpen
  \bibfield  {author} {\bibinfo {author} {\bibfnamefont {P.}~\bibnamefont
  {Campagne-Ibarcq}}, \bibinfo {author} {\bibfnamefont {P.}~\bibnamefont
  {Six}}, \bibinfo {author} {\bibfnamefont {L.}~\bibnamefont {Bretheau}},
  \bibinfo {author} {\bibfnamefont {A.}~\bibnamefont {Sarlette}}, \bibinfo
  {author} {\bibfnamefont {M.}~\bibnamefont {Mirrahimi}}, \bibinfo {author}
  {\bibfnamefont {P.}~\bibnamefont {Rouchon}},\ and\ \bibinfo {author}
  {\bibfnamefont {B.}~\bibnamefont {Huard}},\ }\bibfield  {title} {\bibinfo
  {title} {Observing {Quantum} {State} {Diffusion} by {Heterodyne} {Detection}
  of {Fluorescence}},\ }\href {https://doi.org/10.1103/PhysRevX.6.011002}
  {\bibfield  {journal} {\bibinfo  {journal} {Physical Review X}\ }\textbf
  {\bibinfo {volume} {6}},\ \bibinfo {pages} {011002} (\bibinfo {year}
  {2016})}\BibitemShut {NoStop}%
\bibitem [{\citenamefont {Jordan}\ \emph {et~al.}(2016)\citenamefont {Jordan},
  \citenamefont {Chantasri}, \citenamefont {Rouchon},\ and\ \citenamefont
  {Huard}}]{jordan_anatomy_2016}%
  \BibitemOpen
  \bibfield  {author} {\bibinfo {author} {\bibfnamefont {A.~N.}\ \bibnamefont
  {Jordan}}, \bibinfo {author} {\bibfnamefont {A.}~\bibnamefont {Chantasri}},
  \bibinfo {author} {\bibfnamefont {P.}~\bibnamefont {Rouchon}},\ and\ \bibinfo
  {author} {\bibfnamefont {B.}~\bibnamefont {Huard}},\ }\bibfield  {title}
  {\bibinfo {title} {Anatomy of fluorescence: quantum trajectory statistics
  from continuously measuring spontaneous emission},\ }\href
  {https://doi.org/10.1007/s40509-016-0075-9} {\bibfield  {journal} {\bibinfo
  {journal} {Quantum Studies: Mathematics and Foundations}\ }\textbf {\bibinfo
  {volume} {3}},\ \bibinfo {pages} {237} (\bibinfo {year} {2016})}\BibitemShut
  {NoStop}%
\bibitem [{\citenamefont {Naghiloo}\ \emph {et~al.}(2016)\citenamefont
  {Naghiloo}, \citenamefont {Foroozani}, \citenamefont {Tan}, \citenamefont
  {Jadbabaie},\ and\ \citenamefont {Murch}}]{naghiloo_mapping_2016}%
  \BibitemOpen
  \bibfield  {author} {\bibinfo {author} {\bibfnamefont {M.}~\bibnamefont
  {Naghiloo}}, \bibinfo {author} {\bibfnamefont {N.}~\bibnamefont {Foroozani}},
  \bibinfo {author} {\bibfnamefont {D.}~\bibnamefont {Tan}}, \bibinfo {author}
  {\bibfnamefont {A.}~\bibnamefont {Jadbabaie}},\ and\ \bibinfo {author}
  {\bibfnamefont {K.~W.}\ \bibnamefont {Murch}},\ }\bibfield  {title} {\bibinfo
  {title} {Mapping quantum state dynamics in spontaneous emission},\ }\href
  {https://doi.org/10.1038/ncomms11527} {\bibfield  {journal} {\bibinfo
  {journal} {Nature Communications}\ }\textbf {\bibinfo {volume} {7}},\
  \bibinfo {pages} {11527} (\bibinfo {year} {2016})}\BibitemShut {NoStop}%
\bibitem [{\citenamefont {Ficheux}\ \emph {et~al.}(2018)\citenamefont
  {Ficheux}, \citenamefont {Jezouin}, \citenamefont {Leghtas},\ and\
  \citenamefont {Huard}}]{ficheux_dynamics_2018}%
  \BibitemOpen
  \bibfield  {author} {\bibinfo {author} {\bibfnamefont {Q.}~\bibnamefont
  {Ficheux}}, \bibinfo {author} {\bibfnamefont {S.}~\bibnamefont {Jezouin}},
  \bibinfo {author} {\bibfnamefont {Z.}~\bibnamefont {Leghtas}},\ and\ \bibinfo
  {author} {\bibfnamefont {B.}~\bibnamefont {Huard}},\ }\bibfield  {title}
  {\bibinfo {title} {Dynamics of a qubit while simultaneously monitoring its
  relaxation and dephasing},\ }\href
  {https://doi.org/10.1038/s41467-018-04372-9} {\bibfield  {journal} {\bibinfo
  {journal} {Nature Communications}\ }\textbf {\bibinfo {volume} {9}},\
  \bibinfo {pages} {1926} (\bibinfo {year} {2018})}\BibitemShut {NoStop}%
\bibitem [{\citenamefont {Scigliuzzo}\ \emph {et~al.}(2020)\citenamefont
  {Scigliuzzo}, \citenamefont {Bengtsson}, \citenamefont {Besse}, \citenamefont
  {Wallraff}, \citenamefont {Delsing},\ and\ \citenamefont
  {Gasparinetti}}]{Scigliuzzo2020}%
  \BibitemOpen
  \bibfield  {author} {\bibinfo {author} {\bibfnamefont {M.}~\bibnamefont
  {Scigliuzzo}}, \bibinfo {author} {\bibfnamefont {A.}~\bibnamefont
  {Bengtsson}}, \bibinfo {author} {\bibfnamefont {J.-C.}\ \bibnamefont
  {Besse}}, \bibinfo {author} {\bibfnamefont {A.}~\bibnamefont {Wallraff}},
  \bibinfo {author} {\bibfnamefont {P.}~\bibnamefont {Delsing}},\ and\ \bibinfo
  {author} {\bibfnamefont {S.}~\bibnamefont {Gasparinetti}},\ }\bibfield
  {title} {\bibinfo {title} {{Primary Thermometry of Propagating Microwaves in
  the Quantum Regime}},\ }\href {https://doi.org/10.1103/PhysRevX.10.041054}
  {\bibfield  {journal} {\bibinfo  {journal} {Physical Review X}\ }\textbf
  {\bibinfo {volume} {10}},\ \bibinfo {pages} {041054} (\bibinfo {year}
  {2020})}\BibitemShut {NoStop}%
\bibitem [{\citenamefont {Campagne-Ibarcq}\ \emph {et~al.}(2014)\citenamefont
  {Campagne-Ibarcq}, \citenamefont {Bretheau}, \citenamefont {Flurin},
  \citenamefont {Auffèves}, \citenamefont {Mallet},\ and\ \citenamefont
  {Huard}}]{campagne-ibarcq_observing_2014}%
  \BibitemOpen
  \bibfield  {author} {\bibinfo {author} {\bibfnamefont {P.}~\bibnamefont
  {Campagne-Ibarcq}}, \bibinfo {author} {\bibfnamefont {L.}~\bibnamefont
  {Bretheau}}, \bibinfo {author} {\bibfnamefont {E.}~\bibnamefont {Flurin}},
  \bibinfo {author} {\bibfnamefont {A.}~\bibnamefont {Auffèves}}, \bibinfo
  {author} {\bibfnamefont {F.}~\bibnamefont {Mallet}},\ and\ \bibinfo {author}
  {\bibfnamefont {B.}~\bibnamefont {Huard}},\ }\bibfield  {title} {\bibinfo
  {title} {Observing {Interferences} between {Past} and {Future} {Quantum}
  {States} in {Resonance} {Fluorescence}},\ }\href
  {https://doi.org/10.1103/PhysRevLett.112.180402} {\bibfield  {journal}
  {\bibinfo  {journal} {Physical Review Letters}\ }\textbf {\bibinfo {volume}
  {112}},\ \bibinfo {pages} {180402} (\bibinfo {year} {2014})}\BibitemShut
  {NoStop}%
\bibitem [{\citenamefont {Tan}\ \emph {et~al.}(2017)\citenamefont {Tan},
  \citenamefont {Foroozani}, \citenamefont {Naghiloo}, \citenamefont
  {Kiilerich}, \citenamefont {Mølmer},\ and\ \citenamefont
  {Murch}}]{tan_homodyne_2017}%
  \BibitemOpen
  \bibfield  {author} {\bibinfo {author} {\bibfnamefont {D.}~\bibnamefont
  {Tan}}, \bibinfo {author} {\bibfnamefont {N.}~\bibnamefont {Foroozani}},
  \bibinfo {author} {\bibfnamefont {M.}~\bibnamefont {Naghiloo}}, \bibinfo
  {author} {\bibfnamefont {A.~H.}\ \bibnamefont {Kiilerich}}, \bibinfo {author}
  {\bibfnamefont {K.}~\bibnamefont {Mølmer}},\ and\ \bibinfo {author}
  {\bibfnamefont {K.~W.}\ \bibnamefont {Murch}},\ }\bibfield  {title} {\bibinfo
  {title} {Homodyne monitoring of postselected decay},\ }\href
  {https://doi.org/10.1103/PhysRevA.96.022104} {\bibfield  {journal} {\bibinfo
  {journal} {Physical Review A}\ }\textbf {\bibinfo {volume} {96}},\ \bibinfo
  {pages} {022104} (\bibinfo {year} {2017})}\BibitemShut {NoStop}%
\bibitem [{\citenamefont {Naghiloo}\ \emph {et~al.}(2017)\citenamefont
  {Naghiloo}, \citenamefont {Tan}, \citenamefont {Harrington}, \citenamefont
  {Lewalle}, \citenamefont {Jordan},\ and\ \citenamefont
  {Murch}}]{naghiloo_quantum_2017}%
  \BibitemOpen
  \bibfield  {author} {\bibinfo {author} {\bibfnamefont {M.}~\bibnamefont
  {Naghiloo}}, \bibinfo {author} {\bibfnamefont {D.}~\bibnamefont {Tan}},
  \bibinfo {author} {\bibfnamefont {P.~M.}\ \bibnamefont {Harrington}},
  \bibinfo {author} {\bibfnamefont {P.}~\bibnamefont {Lewalle}}, \bibinfo
  {author} {\bibfnamefont {A.~N.}\ \bibnamefont {Jordan}},\ and\ \bibinfo
  {author} {\bibfnamefont {K.~W.}\ \bibnamefont {Murch}},\ }\bibfield  {title}
  {\bibinfo {title} {Quantum caustics in resonance-fluorescence trajectories},\
  }\href {https://doi.org/10.1103/PhysRevA.96.053807} {\bibfield  {journal}
  {\bibinfo  {journal} {Physical Review A}\ }\textbf {\bibinfo {volume} {96}},\
  \bibinfo {pages} {053807} (\bibinfo {year} {2017})}\BibitemShut {NoStop}%
\bibitem [{noa()}]{noauthor_supplementary_nodate}%
  \BibitemOpen
  \href@noop {} {\bibinfo {title} {See supplemental material [url], which
  includes
  \cite{cottet_electron_2020,pedregosa_scikit-learn_2011,lindblad_generators_1976,haroche_exploring_2006}.}}\BibitemShut
  {Stop}%
\bibitem [{\citenamefont {Cottet}\ \emph {et~al.}(2021)\citenamefont {Cottet},
  \citenamefont {Xiong}, \citenamefont {Nguyen}, \citenamefont {Lin},\ and\
  \citenamefont {Manucharyan}}]{cottet_electron_2020}%
  \BibitemOpen
  \bibfield  {author} {\bibinfo {author} {\bibfnamefont {N.}~\bibnamefont
  {Cottet}}, \bibinfo {author} {\bibfnamefont {H.}~\bibnamefont {Xiong}},
  \bibinfo {author} {\bibfnamefont {L.~B.}\ \bibnamefont {Nguyen}}, \bibinfo
  {author} {\bibfnamefont {Y.-H.}\ \bibnamefont {Lin}},\ and\ \bibinfo {author}
  {\bibfnamefont {V.~E.}\ \bibnamefont {Manucharyan}},\ }\bibfield  {title}
  {\bibinfo {title} {{Electron shelving of a superconducting artificial
  atom}},\ }\href {https://doi.org/10.1038/s41467-021-26686-x} {\bibfield
  {journal} {\bibinfo  {journal} {Nature Communications}\ }\textbf {\bibinfo
  {volume} {12}},\ \bibinfo {pages} {6383} (\bibinfo {year}
  {2021})}\BibitemShut {NoStop}%
\bibitem [{\citenamefont {Pedregosa}\ \emph {et~al.}(2011)\citenamefont
  {Pedregosa}, \citenamefont {Varoquaux}, \citenamefont {Gramfort},
  \citenamefont {Michel}, \citenamefont {Thirion}, \citenamefont {Grisel},
  \citenamefont {Blondel}, \citenamefont {Prettenhofer}, \citenamefont {Weiss},
  \citenamefont {Dubourg}, \citenamefont {Vanderplas}, \citenamefont {Passos},
  \citenamefont {Cournapeau}, \citenamefont {Brucher}, \citenamefont {Perrot},\
  and\ \citenamefont {Duchesnay}}]{pedregosa_scikit-learn_2011}%
  \BibitemOpen
  \bibfield  {author} {\bibinfo {author} {\bibfnamefont {F.}~\bibnamefont
  {Pedregosa}}, \bibinfo {author} {\bibfnamefont {G.}~\bibnamefont
  {Varoquaux}}, \bibinfo {author} {\bibfnamefont {A.}~\bibnamefont {Gramfort}},
  \bibinfo {author} {\bibfnamefont {V.}~\bibnamefont {Michel}}, \bibinfo
  {author} {\bibfnamefont {B.}~\bibnamefont {Thirion}}, \bibinfo {author}
  {\bibfnamefont {O.}~\bibnamefont {Grisel}}, \bibinfo {author} {\bibfnamefont
  {M.}~\bibnamefont {Blondel}}, \bibinfo {author} {\bibfnamefont
  {P.}~\bibnamefont {Prettenhofer}}, \bibinfo {author} {\bibfnamefont
  {R.}~\bibnamefont {Weiss}}, \bibinfo {author} {\bibfnamefont
  {V.}~\bibnamefont {Dubourg}}, \bibinfo {author} {\bibfnamefont
  {J.}~\bibnamefont {Vanderplas}}, \bibinfo {author} {\bibfnamefont
  {A.}~\bibnamefont {Passos}}, \bibinfo {author} {\bibfnamefont
  {D.}~\bibnamefont {Cournapeau}}, \bibinfo {author} {\bibfnamefont
  {M.}~\bibnamefont {Brucher}}, \bibinfo {author} {\bibfnamefont
  {M.}~\bibnamefont {Perrot}},\ and\ \bibinfo {author} {\bibfnamefont
  {E.}~\bibnamefont {Duchesnay}},\ }\bibfield  {title} {\bibinfo {title}
  {Scikit-learn: {Machine} {Learning} in {Python}},\ }\href@noop {} {\bibfield
  {journal} {\bibinfo  {journal} {Journal of Machine Learning Research}\
  }\textbf {\bibinfo {volume} {12}},\ \bibinfo {pages} {2825} (\bibinfo {year}
  {2011})}\BibitemShut {NoStop}%
\bibitem [{\citenamefont {Lindblad}(1976)}]{lindblad_generators_1976}%
  \BibitemOpen
  \bibfield  {author} {\bibinfo {author} {\bibfnamefont {G.}~\bibnamefont
  {Lindblad}},\ }\bibfield  {title} {\bibinfo {title} {On the generators of
  quantum dynamical semigroups},\ }\href {https://doi.org/10.1007/BF01608499}
  {\bibfield  {journal} {\bibinfo  {journal} {Communications in Mathematical
  Physics}\ }\textbf {\bibinfo {volume} {48}},\ \bibinfo {pages} {119}
  (\bibinfo {year} {1976})}\BibitemShut {NoStop}%
\bibitem [{\citenamefont {Haroche}\ and\ \citenamefont
  {Raimond}(2006)}]{haroche_exploring_2006}%
  \BibitemOpen
  \bibfield  {author} {\bibinfo {author} {\bibfnamefont {S.}~\bibnamefont
  {Haroche}}\ and\ \bibinfo {author} {\bibfnamefont {J.-M.}\ \bibnamefont
  {Raimond}},\ }\href@noop {} {\emph {\bibinfo {title} {Exploring the
  {Quantum}. {Atoms}, {Cavities} and {Photons}.}}},\ \bibinfo {edition} {oxford
  graduate texts}\ ed.\ (\bibinfo  {publisher} {Oxford University Press},\
  \bibinfo {year} {2006})\BibitemShut {NoStop}%
\bibitem [{\citenamefont {Van~Enk}\ and\ \citenamefont
  {Kimble}(2002)}]{vanEnk2002}%
  \BibitemOpen
  \bibfield  {author} {\bibinfo {author} {\bibfnamefont {S.~J.}\ \bibnamefont
  {Van~Enk}}\ and\ \bibinfo {author} {\bibfnamefont {H.~J.}\ \bibnamefont
  {Kimble}},\ }\bibfield  {title} {\bibinfo {title} {On the classical character
  of control fields in quantum information processing},\ }\href@noop {}
  {\bibfield  {journal} {\bibinfo  {journal} {Quantum Info. Comput.}\ }\textbf
  {\bibinfo {volume} {2}},\ \bibinfo {pages} {1–13} (\bibinfo {year}
  {2002})}\BibitemShut {NoStop}%
\bibitem [{\citenamefont {Silberfarb}\ and\ \citenamefont
  {Deutsch}(2004)}]{Silberfarb2004}%
  \BibitemOpen
  \bibfield  {author} {\bibinfo {author} {\bibfnamefont {A.}~\bibnamefont
  {Silberfarb}}\ and\ \bibinfo {author} {\bibfnamefont {I.~H.}\ \bibnamefont
  {Deutsch}},\ }\bibfield  {title} {\bibinfo {title} {{Entanglement generated
  between a single atom and a laser pulse}},\ }\href
  {https://doi.org/10.1103/PhysRevA.69.042308} {\bibfield  {journal} {\bibinfo
  {journal} {Physical Review A}\ }\textbf {\bibinfo {volume} {69}},\ \bibinfo
  {pages} {042308} (\bibinfo {year} {2004})}\BibitemShut {NoStop}%
\bibitem [{\citenamefont {Itano}(2003)}]{Itano2003}%
  \BibitemOpen
  \bibfield  {author} {\bibinfo {author} {\bibfnamefont {W.~M.}\ \bibnamefont
  {Itano}},\ }\bibfield  {title} {\bibinfo {title} {{Comment on “Some
  implications of the quantum nature of laser fields for quantum
  computations”}},\ }\href {https://doi.org/10.1103/PhysRevA.68.046301}
  {\bibfield  {journal} {\bibinfo  {journal} {Physical Review A}\ }\textbf
  {\bibinfo {volume} {68}},\ \bibinfo {pages} {046301} (\bibinfo {year}
  {2003})}\BibitemShut {NoStop}%
\bibitem [{\citenamefont {Silberfarb}\ and\ \citenamefont
  {Deutsch}(2003)}]{Silberfarb2003}%
  \BibitemOpen
  \bibfield  {author} {\bibinfo {author} {\bibfnamefont {A.}~\bibnamefont
  {Silberfarb}}\ and\ \bibinfo {author} {\bibfnamefont {I.~H.}\ \bibnamefont
  {Deutsch}},\ }\bibfield  {title} {\bibinfo {title} {{Continuous measurement
  with traveling-wave probes}},\ }\href
  {https://doi.org/10.1103/PhysRevA.68.013817} {\bibfield  {journal} {\bibinfo
  {journal} {Physical Review A}\ }\textbf {\bibinfo {volume} {68}},\ \bibinfo
  {pages} {013817} (\bibinfo {year} {2003})}\BibitemShut {NoStop}%
\bibitem [{\citenamefont {Nha}\ and\ \citenamefont
  {Carmichael}(2005)}]{Nha2005}%
  \BibitemOpen
  \bibfield  {author} {\bibinfo {author} {\bibfnamefont {H.}~\bibnamefont
  {Nha}}\ and\ \bibinfo {author} {\bibfnamefont {H.~J.}\ \bibnamefont
  {Carmichael}},\ }\bibfield  {title} {\bibinfo {title} {{Decoherence of a
  two-state atom driven by coherent light}},\ }\href
  {https://doi.org/10.1103/PhysRevA.71.013805} {\bibfield  {journal} {\bibinfo
  {journal} {Physical Review A}\ }\textbf {\bibinfo {volume} {71}},\ \bibinfo
  {pages} {013805} (\bibinfo {year} {2005})}\BibitemShut {NoStop}%
\bibitem [{\citenamefont {Gea-Banacloche}\ and\ \citenamefont
  {Ozawa}(2005)}]{Gea_Banacloche_2005}%
  \BibitemOpen
  \bibfield  {author} {\bibinfo {author} {\bibfnamefont {J.}~\bibnamefont
  {Gea-Banacloche}}\ and\ \bibinfo {author} {\bibfnamefont {M.}~\bibnamefont
  {Ozawa}},\ }\bibfield  {title} {\bibinfo {title} {{Constraints for quantum
  logic arising from conservation laws and field fluctuations}},\ }\href
  {https://doi.org/10.1088/1464-4266/7/10/017} {\bibfield  {journal} {\bibinfo
  {journal} {Journal of Optics B: Quantum and Semiclassical Optics}\ }\textbf
  {\bibinfo {volume} {7}},\ \bibinfo {pages} {S326} (\bibinfo {year}
  {2005})}\BibitemShut {NoStop}%
\bibitem [{\citenamefont {Bertet}\ \emph {et~al.}(2001)\citenamefont {Bertet},
  \citenamefont {Osnaghi}, \citenamefont {Rauschenbeutel}, \citenamefont
  {Nogues}, \citenamefont {Auffeves}, \citenamefont {Brune}, \citenamefont
  {Raimond},\ and\ \citenamefont {Haroche}}]{bertet_complementarity_2001}%
  \BibitemOpen
  \bibfield  {author} {\bibinfo {author} {\bibfnamefont {P.}~\bibnamefont
  {Bertet}}, \bibinfo {author} {\bibfnamefont {S.}~\bibnamefont {Osnaghi}},
  \bibinfo {author} {\bibfnamefont {A.}~\bibnamefont {Rauschenbeutel}},
  \bibinfo {author} {\bibfnamefont {G.}~\bibnamefont {Nogues}}, \bibinfo
  {author} {\bibfnamefont {A.}~\bibnamefont {Auffeves}}, \bibinfo {author}
  {\bibfnamefont {M.}~\bibnamefont {Brune}}, \bibinfo {author} {\bibfnamefont
  {J.~M.}\ \bibnamefont {Raimond}},\ and\ \bibinfo {author} {\bibfnamefont
  {S.}~\bibnamefont {Haroche}},\ }\bibfield  {title} {\bibinfo {title} {A
  complementarity experiment with an interferometer at the quantum–classical
  boundary},\ }\href {https://doi.org/10.1038/35075517} {\bibfield  {journal}
  {\bibinfo  {journal} {Nature}\ }\textbf {\bibinfo {volume} {411}},\ \bibinfo
  {pages} {166} (\bibinfo {year} {2001})}\BibitemShut {NoStop}%
\bibitem [{\citenamefont {Macklin}\ \emph {et~al.}(2015)\citenamefont
  {Macklin}, \citenamefont {O’Brien}, \citenamefont {Hover}, \citenamefont
  {Schwartz}, \citenamefont {Bolkhovsky}, \citenamefont {Zhang}, \citenamefont
  {Oliver},\ and\ \citenamefont {Siddiqi}}]{macklin_nearquantum-limited_2015}%
  \BibitemOpen
  \bibfield  {author} {\bibinfo {author} {\bibfnamefont {C.}~\bibnamefont
  {Macklin}}, \bibinfo {author} {\bibfnamefont {K.}~\bibnamefont {O’Brien}},
  \bibinfo {author} {\bibfnamefont {D.}~\bibnamefont {Hover}}, \bibinfo
  {author} {\bibfnamefont {M.~E.}\ \bibnamefont {Schwartz}}, \bibinfo {author}
  {\bibfnamefont {V.}~\bibnamefont {Bolkhovsky}}, \bibinfo {author}
  {\bibfnamefont {X.}~\bibnamefont {Zhang}}, \bibinfo {author} {\bibfnamefont
  {W.~D.}\ \bibnamefont {Oliver}},\ and\ \bibinfo {author} {\bibfnamefont
  {I.}~\bibnamefont {Siddiqi}},\ }\bibfield  {title} {\bibinfo {title} {A
  near–quantum-limited {Josephson} traveling-wave parametric amplifier},\
  }\href {https://doi.org/10.1126/science.aaa8525} {\bibfield  {journal}
  {\bibinfo  {journal} {Science}\ }\textbf {\bibinfo {volume} {350}},\ \bibinfo
  {pages} {307} (\bibinfo {year} {2015})}\BibitemShut {NoStop}%
\bibitem [{\citenamefont {Vool}\ and\ \citenamefont
  {Devoret}(2017)}]{vool_introduction_2017}%
  \BibitemOpen
  \bibfield  {author} {\bibinfo {author} {\bibfnamefont {U.}~\bibnamefont
  {Vool}}\ and\ \bibinfo {author} {\bibfnamefont {M.}~\bibnamefont {Devoret}},\
  }\bibfield  {title} {\bibinfo {title} {Introduction to quantum
  electromagnetic circuits},\ }\href
  {https://doi.org/https://doi.org/10.1002/cta.2359} {\bibfield  {journal}
  {\bibinfo  {journal} {International Journal of Circuit Theory and
  Applications}\ }\textbf {\bibinfo {volume} {45}},\ \bibinfo {pages} {897}
  (\bibinfo {year} {2017})}\BibitemShut {NoStop}%
\bibitem [{\citenamefont {Clerk}\ \emph {et~al.}(2010)\citenamefont {Clerk},
  \citenamefont {Devoret}, \citenamefont {Girvin}, \citenamefont {Marquardt},\
  and\ \citenamefont {Schoelkopf}}]{clerk_introduction_2010}%
  \BibitemOpen
  \bibfield  {author} {\bibinfo {author} {\bibfnamefont {A.~A.}\ \bibnamefont
  {Clerk}}, \bibinfo {author} {\bibfnamefont {M.~H.}\ \bibnamefont {Devoret}},
  \bibinfo {author} {\bibfnamefont {S.~M.}\ \bibnamefont {Girvin}}, \bibinfo
  {author} {\bibfnamefont {F.}~\bibnamefont {Marquardt}},\ and\ \bibinfo
  {author} {\bibfnamefont {R.~J.}\ \bibnamefont {Schoelkopf}},\ }\bibfield
  {title} {\bibinfo {title} {Introduction to quantum noise, measurement, and
  amplification},\ }\href {https://doi.org/10.1103/RevModPhys.82.1155}
  {\bibfield  {journal} {\bibinfo  {journal} {Reviews of Modern Physics}\
  }\textbf {\bibinfo {volume} {82}},\ \bibinfo {pages} {1155} (\bibinfo {year}
  {2010})}\BibitemShut {NoStop}%
\bibitem [{\citenamefont {Cottet}(2018)}]{cottet_energy_2018}%
  \BibitemOpen
  \bibfield  {author} {\bibinfo {author} {\bibfnamefont {N.}~\bibnamefont
  {Cottet}},\ }\emph {\bibinfo {title} {Energy and information in fluorescence
  with superconducting circuits}},\ \href@noop {} {\bibinfo {type} {{PhD}
  {Thesis}}},\ \bibinfo  {school} {ENS Paris} (\bibinfo {year}
  {2018})\BibitemShut {NoStop}%
\bibitem [{\citenamefont {Wiseman}(2002)}]{wiseman_weak_2002}%
  \BibitemOpen
  \bibfield  {author} {\bibinfo {author} {\bibfnamefont {H.~M.}\ \bibnamefont
  {Wiseman}},\ }\bibfield  {title} {\bibinfo {title} {Weak values, quantum
  trajectories, and the cavity-{QED} experiment on wave-particle correlation},\
  }\href {https://doi.org/10.1103/PhysRevA.65.032111} {\bibfield  {journal}
  {\bibinfo  {journal} {Physical Review A}\ }\textbf {\bibinfo {volume} {65}},\
  \bibinfo {pages} {032111} (\bibinfo {year} {2002})}\BibitemShut {NoStop}%
\bibitem [{\citenamefont {Tsang}(2009)}]{tsang_optimal_2009}%
  \BibitemOpen
  \bibfield  {author} {\bibinfo {author} {\bibfnamefont {M.}~\bibnamefont
  {Tsang}},\ }\bibfield  {title} {\bibinfo {title} {Optimal waveform estimation
  for classical and quantum systems via time-symmetric smoothing},\ }\href
  {https://doi.org/10.1103/PhysRevA.80.033840} {\bibfield  {journal} {\bibinfo
  {journal} {Physical Review A}\ }\textbf {\bibinfo {volume} {80}},\ \bibinfo
  {pages} {033840} (\bibinfo {year} {2009})}\BibitemShut {NoStop}%
\bibitem [{\citenamefont {Gammelmark}\ \emph {et~al.}(2013)\citenamefont
  {Gammelmark}, \citenamefont {Julsgaard},\ and\ \citenamefont
  {Molmer}}]{gammelmark_past_2013}%
  \BibitemOpen
  \bibfield  {author} {\bibinfo {author} {\bibfnamefont {S.}~\bibnamefont
  {Gammelmark}}, \bibinfo {author} {\bibfnamefont {B.}~\bibnamefont
  {Julsgaard}},\ and\ \bibinfo {author} {\bibfnamefont {K.}~\bibnamefont
  {Molmer}},\ }\bibfield  {title} {\bibinfo {title} {Past {Quantum} {States} of
  a {Monitored} {System}},\ }\href
  {https://doi.org/10.1103/PhysRevLett.111.160401} {\bibfield  {journal}
  {\bibinfo  {journal} {Physical Review Letters}\ }\textbf {\bibinfo {volume}
  {111}},\ \bibinfo {pages} {160401} (\bibinfo {year} {2013})}\BibitemShut
  {NoStop}%
\bibitem [{\citenamefont {Maffei}\ \emph {et~al.}(2021)\citenamefont {Maffei},
  \citenamefont {Camati},\ and\ \citenamefont
  {Auff{\`{e}}ves}}]{maffei_probing_2021}%
  \BibitemOpen
  \bibfield  {author} {\bibinfo {author} {\bibfnamefont {M.}~\bibnamefont
  {Maffei}}, \bibinfo {author} {\bibfnamefont {P.~A.}\ \bibnamefont {Camati}},\
  and\ \bibinfo {author} {\bibfnamefont {A.}~\bibnamefont {Auff{\`{e}}ves}},\
  }\bibfield  {title} {\bibinfo {title} {{Probing nonclassical light fields
  with energetic witnesses in waveguide quantum electrodynamics}},\ }\href
  {https://doi.org/10.1103/PhysRevResearch.3.L032073} {\bibfield  {journal}
  {\bibinfo  {journal} {Physical Review Research}\ }\textbf {\bibinfo {volume}
  {3}},\ \bibinfo {pages} {L032073} (\bibinfo {year} {2021})}\BibitemShut
  {NoStop}%
\bibitem [{\citenamefont {Maffei~et. al.}()}]{maffei_manuscript_nodate}%
  \BibitemOpen
  \bibfield  {author} {\bibinfo {author} {\bibfnamefont {M.}~\bibnamefont
  {Maffei~et. al.}},\ }\href@noop {} {\bibinfo {title} {In
  preparation.}}\BibitemShut {Stop}%
\bibitem [{\citenamefont {Mizrahi}\ and\ \citenamefont
  {Dodonov}(2002)}]{Mizrahi2002}%
  \BibitemOpen
  \bibfield  {author} {\bibinfo {author} {\bibfnamefont {S.~S.}\ \bibnamefont
  {Mizrahi}}\ and\ \bibinfo {author} {\bibfnamefont {V.~V.}\ \bibnamefont
  {Dodonov}},\ }\bibfield  {title} {\bibinfo {title} {{Creating quanta with an
  annihilation operator}},\ }\href
  {https://doi.org/10.1088/0305-4470/35/41/315} {\bibfield  {journal} {\bibinfo
   {journal} {Journal of Physics A: Mathematical and General}\ }\textbf
  {\bibinfo {volume} {35}},\ \bibinfo {pages} {8847} (\bibinfo {year}
  {2002})}\BibitemShut {NoStop}%
\bibitem [{\citenamefont {Dressel}\ \emph {et~al.}(2014)\citenamefont
  {Dressel}, \citenamefont {Malik}, \citenamefont {Miatto}, \citenamefont
  {Jordan},\ and\ \citenamefont {Boyd}}]{dressel_colloquium_2014}%
  \BibitemOpen
  \bibfield  {author} {\bibinfo {author} {\bibfnamefont {J.}~\bibnamefont
  {Dressel}}, \bibinfo {author} {\bibfnamefont {M.}~\bibnamefont {Malik}},
  \bibinfo {author} {\bibfnamefont {F.~M.}\ \bibnamefont {Miatto}}, \bibinfo
  {author} {\bibfnamefont {A.~N.}\ \bibnamefont {Jordan}},\ and\ \bibinfo
  {author} {\bibfnamefont {R.~W.}\ \bibnamefont {Boyd}},\ }\bibfield  {title}
  {\bibinfo {title} {Colloquium: {Understanding} quantum weak values: {Basics}
  and applications},\ }\href {https://doi.org/10.1103/RevModPhys.86.307}
  {\bibfield  {journal} {\bibinfo  {journal} {Reviews of Modern Physics}\
  }\textbf {\bibinfo {volume} {86}},\ \bibinfo {pages} {307} (\bibinfo {year}
  {2014})}\BibitemShut {NoStop}%
\bibitem [{\citenamefont {Ghosh}\ and\ \citenamefont {Gerry}(1997)}]{Ghosh:97}%
  \BibitemOpen
  \bibfield  {author} {\bibinfo {author} {\bibfnamefont {H.}~\bibnamefont
  {Ghosh}}\ and\ \bibinfo {author} {\bibfnamefont {C.~C.}\ \bibnamefont
  {Gerry}},\ }\bibfield  {title} {\bibinfo {title} {Measurement-induced
  nonclassical states of the {Jaynes}--{Cummings} model},\ }\href
  {https://doi.org/10.1364/JOSAB.14.002782} {\bibfield  {journal} {\bibinfo
  {journal} {J. Opt. Soc. Am. B}\ }\textbf {\bibinfo {volume} {14}},\ \bibinfo
  {pages} {2782} (\bibinfo {year} {1997})}\BibitemShut {NoStop}%
\bibitem [{\citenamefont {Fischer}\ \emph {et~al.}(2018)\citenamefont
  {Fischer}, \citenamefont {Trivedi}, \citenamefont {Ramasesh}, \citenamefont
  {Siddiqi},\ and\ \citenamefont {Vučković}}]{fischer_scattering_2018}%
  \BibitemOpen
  \bibfield  {author} {\bibinfo {author} {\bibfnamefont {K.~A.}\ \bibnamefont
  {Fischer}}, \bibinfo {author} {\bibfnamefont {R.}~\bibnamefont {Trivedi}},
  \bibinfo {author} {\bibfnamefont {V.}~\bibnamefont {Ramasesh}}, \bibinfo
  {author} {\bibfnamefont {I.}~\bibnamefont {Siddiqi}},\ and\ \bibinfo {author}
  {\bibfnamefont {J.}~\bibnamefont {Vučković}},\ }\bibfield  {title}
  {\bibinfo {title} {Scattering into one-dimensional waveguides from a
  coherently-driven quantum-optical system},\ }\href
  {https://doi.org/10.22331/q-2018-05-28-69} {\bibfield  {journal} {\bibinfo
  {journal} {Quantum}\ }\textbf {\bibinfo {volume} {2}},\ \bibinfo {pages} {69}
  (\bibinfo {year} {2018})}\BibitemShut {NoStop}%
\bibitem [{\citenamefont {Ueda}\ \emph {et~al.}(1992)\citenamefont {Ueda},
  \citenamefont {Imoto}, \citenamefont {Nagaoka},\ and\ \citenamefont
  {Ogawa}}]{Ueda1992}%
  \BibitemOpen
  \bibfield  {author} {\bibinfo {author} {\bibfnamefont {M.}~\bibnamefont
  {Ueda}}, \bibinfo {author} {\bibfnamefont {N.}~\bibnamefont {Imoto}},
  \bibinfo {author} {\bibfnamefont {H.}~\bibnamefont {Nagaoka}},\ and\ \bibinfo
  {author} {\bibfnamefont {T.}~\bibnamefont {Ogawa}},\ }\bibfield  {title}
  {\bibinfo {title} {{Continuous quantum-nondemolition measurement of photon
  number}},\ }\href {https://doi.org/10.1103/PhysRevA.46.2859} {\bibfield
  {journal} {\bibinfo  {journal} {Physical Review A}\ }\textbf {\bibinfo
  {volume} {46}},\ \bibinfo {pages} {2859} (\bibinfo {year}
  {1992})}\BibitemShut {NoStop}%
\bibitem [{\citenamefont {Nunn}\ \emph {et~al.}(2022)\citenamefont {Nunn},
  \citenamefont {Franson},\ and\ \citenamefont {Pittman}}]{Nunn2022}%
  \BibitemOpen
  \bibfield  {author} {\bibinfo {author} {\bibfnamefont {C.~M.}\ \bibnamefont
  {Nunn}}, \bibinfo {author} {\bibfnamefont {J.~D.}\ \bibnamefont {Franson}},\
  and\ \bibinfo {author} {\bibfnamefont {T.~B.}\ \bibnamefont {Pittman}},\
  }\bibfield  {title} {\bibinfo {title} {{Modifying quantum optical states by
  zero-photon subtraction}},\ }\href
  {https://doi.org/10.1103/PhysRevA.105.033702} {\bibfield  {journal} {\bibinfo
   {journal} {Physical Review A}\ }\textbf {\bibinfo {volume} {105}},\ \bibinfo
  {pages} {033702} (\bibinfo {year} {2022})}\BibitemShut {NoStop}%
\bibitem [{\citenamefont {Besse}\ \emph {et~al.}(2020)\citenamefont {Besse},
  \citenamefont {Gasparinetti}, \citenamefont {Collodo}, \citenamefont
  {Walter}, \citenamefont {Remm}, \citenamefont {Krause}, \citenamefont
  {Eichler},\ and\ \citenamefont {Wallraff}}]{besse_parity_2020}%
  \BibitemOpen
  \bibfield  {author} {\bibinfo {author} {\bibfnamefont {J.-C.}\ \bibnamefont
  {Besse}}, \bibinfo {author} {\bibfnamefont {S.}~\bibnamefont {Gasparinetti}},
  \bibinfo {author} {\bibfnamefont {M.~C.}\ \bibnamefont {Collodo}}, \bibinfo
  {author} {\bibfnamefont {T.}~\bibnamefont {Walter}}, \bibinfo {author}
  {\bibfnamefont {A.}~\bibnamefont {Remm}}, \bibinfo {author} {\bibfnamefont
  {J.}~\bibnamefont {Krause}}, \bibinfo {author} {\bibfnamefont
  {C.}~\bibnamefont {Eichler}},\ and\ \bibinfo {author} {\bibfnamefont
  {A.}~\bibnamefont {Wallraff}},\ }\bibfield  {title} {\bibinfo {title} {Parity
  {Detection} of {Propagating} {Microwave} {Fields}},\ }\href
  {https://doi.org/10.1103/PhysRevX.10.011046} {\bibfield  {journal} {\bibinfo
  {journal} {Physical Review X}\ }\textbf {\bibinfo {volume} {10}},\ \bibinfo
  {pages} {011046} (\bibinfo {year} {2020})}\BibitemShut {NoStop}%
\bibitem [{\citenamefont {Dassonneville}\ \emph {et~al.}(2020)\citenamefont
  {Dassonneville}, \citenamefont {Assouly}, \citenamefont {Peronnin},
  \citenamefont {Rouchon},\ and\ \citenamefont
  {Huard}}]{dassonneville_number-resolved_2020}%
  \BibitemOpen
  \bibfield  {author} {\bibinfo {author} {\bibfnamefont {R.}~\bibnamefont
  {Dassonneville}}, \bibinfo {author} {\bibfnamefont {R.}~\bibnamefont
  {Assouly}}, \bibinfo {author} {\bibfnamefont {T.}~\bibnamefont {Peronnin}},
  \bibinfo {author} {\bibfnamefont {P.}~\bibnamefont {Rouchon}},\ and\ \bibinfo
  {author} {\bibfnamefont {B.}~\bibnamefont {Huard}},\ }\bibfield  {title}
  {\bibinfo {title} {Number-{Resolved} {Photocounter} for {Propagating}
  {Microwave} {Mode}},\ }\href
  {https://doi.org/10.1103/PhysRevApplied.14.044022} {\bibfield  {journal}
  {\bibinfo  {journal} {Physical Review Applied}\ }\textbf {\bibinfo {volume}
  {14}},\ \bibinfo {pages} {044022} (\bibinfo {year} {2020})}\BibitemShut
  {NoStop}%
\bibitem [{\citenamefont {Shahmoon}\ \emph {et~al.}(2009)\citenamefont
  {Shahmoon}, \citenamefont {Levit},\ and\ \citenamefont
  {Ozeri}}]{Shahmoon2009}%
  \BibitemOpen
  \bibfield  {author} {\bibinfo {author} {\bibfnamefont {E.}~\bibnamefont
  {Shahmoon}}, \bibinfo {author} {\bibfnamefont {S.}~\bibnamefont {Levit}},\
  and\ \bibinfo {author} {\bibfnamefont {R.}~\bibnamefont {Ozeri}},\ }\bibfield
   {title} {\bibinfo {title} {{Qubit coherent control and entanglement with
  squeezed light fields}},\ }\href {https://doi.org/10.1103/PhysRevA.80.033803}
  {\bibfield  {journal} {\bibinfo  {journal} {Physical Review A}\ }\textbf
  {\bibinfo {volume} {80}},\ \bibinfo {pages} {033803} (\bibinfo {year}
  {2009})}\BibitemShut {NoStop}%
\bibitem [{\citenamefont {Goldberg}\ and\ \citenamefont
  {Steinberg}(2020)}]{Goldberg2020}%
  \BibitemOpen
  \bibfield  {author} {\bibinfo {author} {\bibfnamefont {A.~Z.}\ \bibnamefont
  {Goldberg}}\ and\ \bibinfo {author} {\bibfnamefont {A.~M.}\ \bibnamefont
  {Steinberg}},\ }\bibfield  {title} {\bibinfo {title} {{Transcoherent States:
  Optical States for Maximal Generation of Atomic Coherence}},\ }\href
  {https://doi.org/10.1103/PRXQuantum.1.020306} {\bibfield  {journal} {\bibinfo
   {journal} {PRX Quantum}\ }\textbf {\bibinfo {volume} {1}},\ \bibinfo {pages}
  {020306} (\bibinfo {year} {2020})}\BibitemShut {NoStop}%
\bibitem [{\citenamefont {Brandner}\ \emph {et~al.}(2015)\citenamefont
  {Brandner}, \citenamefont {Bauer}, \citenamefont {Schmid},\ and\
  \citenamefont {Seifert}}]{brandner_coherence-enhanced_2015}%
  \BibitemOpen
  \bibfield  {author} {\bibinfo {author} {\bibfnamefont {K.}~\bibnamefont
  {Brandner}}, \bibinfo {author} {\bibfnamefont {M.}~\bibnamefont {Bauer}},
  \bibinfo {author} {\bibfnamefont {M.~T.}\ \bibnamefont {Schmid}},\ and\
  \bibinfo {author} {\bibfnamefont {U.}~\bibnamefont {Seifert}},\ }\bibfield
  {title} {\bibinfo {title} {Coherence-enhanced efficiency of feedback-driven
  quantum engines},\ }\href {https://doi.org/10.1088/1367-2630/17/6/065006}
  {\bibfield  {journal} {\bibinfo  {journal} {New Journal of Physics}\ }\textbf
  {\bibinfo {volume} {17}},\ \bibinfo {pages} {065006} (\bibinfo {year}
  {2015})}\BibitemShut {NoStop}%
\bibitem [{\citenamefont {Yi}\ \emph {et~al.}(2017)\citenamefont {Yi},
  \citenamefont {Talkner},\ and\ \citenamefont
  {Kim}}]{yi_single-temperature_2017}%
  \BibitemOpen
  \bibfield  {author} {\bibinfo {author} {\bibfnamefont {J.}~\bibnamefont
  {Yi}}, \bibinfo {author} {\bibfnamefont {P.}~\bibnamefont {Talkner}},\ and\
  \bibinfo {author} {\bibfnamefont {Y.~W.}\ \bibnamefont {Kim}},\ }\bibfield
  {title} {\bibinfo {title} {Single-temperature quantum engine without feedback
  control},\ }\href {https://doi.org/10.1103/PhysRevE.96.022108} {\bibfield
  {journal} {\bibinfo  {journal} {Physical Review E}\ }\textbf {\bibinfo
  {volume} {96}},\ \bibinfo {pages} {022108} (\bibinfo {year}
  {2017})}\BibitemShut {NoStop}%
\bibitem [{\citenamefont {Elouard}\ \emph {et~al.}(2017)\citenamefont
  {Elouard}, \citenamefont {Herrera-Martí}, \citenamefont {Huard},\ and\
  \citenamefont {Auffèves}}]{elouard_extracting_2017}%
  \BibitemOpen
  \bibfield  {author} {\bibinfo {author} {\bibfnamefont {C.}~\bibnamefont
  {Elouard}}, \bibinfo {author} {\bibfnamefont {D.}~\bibnamefont
  {Herrera-Martí}}, \bibinfo {author} {\bibfnamefont {B.}~\bibnamefont
  {Huard}},\ and\ \bibinfo {author} {\bibfnamefont {A.}~\bibnamefont
  {Auffèves}},\ }\bibfield  {title} {\bibinfo {title} {Extracting {Work} from
  {Quantum} {Measurement} in {Maxwell}'s {Demon} {Engines}},\ }\href
  {https://doi.org/10.1103/PhysRevLett.118.260603} {\bibfield  {journal}
  {\bibinfo  {journal} {Physical Review Letters}\ }\textbf {\bibinfo {volume}
  {118}},\ \bibinfo {pages} {260603} (\bibinfo {year} {2017})}\BibitemShut
  {NoStop}%
\bibitem [{\citenamefont {Elouard}\ and\ \citenamefont
  {Jordan}(2018)}]{elouard_efficient_2018}%
  \BibitemOpen
  \bibfield  {author} {\bibinfo {author} {\bibfnamefont {C.}~\bibnamefont
  {Elouard}}\ and\ \bibinfo {author} {\bibfnamefont {A.~N.}\ \bibnamefont
  {Jordan}},\ }\bibfield  {title} {\bibinfo {title} {Efficient {Quantum}
  {Measurement} {Engines}},\ }\href
  {https://doi.org/10.1103/PhysRevLett.120.260601} {\bibfield  {journal}
  {\bibinfo  {journal} {Physical Review Letters}\ }\textbf {\bibinfo {volume}
  {120}},\ \bibinfo {pages} {260601} (\bibinfo {year} {2018})}\BibitemShut
  {NoStop}%
\bibitem [{\citenamefont {Ding}\ \emph {et~al.}(2018)\citenamefont {Ding},
  \citenamefont {Yi}, \citenamefont {Kim},\ and\ \citenamefont
  {Talkner}}]{ding_measurement-driven_2018}%
  \BibitemOpen
  \bibfield  {author} {\bibinfo {author} {\bibfnamefont {X.}~\bibnamefont
  {Ding}}, \bibinfo {author} {\bibfnamefont {J.}~\bibnamefont {Yi}}, \bibinfo
  {author} {\bibfnamefont {Y.~W.}\ \bibnamefont {Kim}},\ and\ \bibinfo {author}
  {\bibfnamefont {P.}~\bibnamefont {Talkner}},\ }\bibfield  {title} {\bibinfo
  {title} {Measurement-driven single temperature engine},\ }\href
  {https://doi.org/10.1103/PhysRevE.98.042122} {\bibfield  {journal} {\bibinfo
  {journal} {Physical Review E}\ }\textbf {\bibinfo {volume} {98}},\ \bibinfo
  {pages} {042122} (\bibinfo {year} {2018})}\BibitemShut {NoStop}%
\bibitem [{\citenamefont {Abdelkhalek}\ \emph {et~al.}(2018)\citenamefont
  {Abdelkhalek}, \citenamefont {Nakata},\ and\ \citenamefont
  {Reeb}}]{abdelkhalek_fundamental_2018}%
  \BibitemOpen
  \bibfield  {author} {\bibinfo {author} {\bibfnamefont {K.}~\bibnamefont
  {Abdelkhalek}}, \bibinfo {author} {\bibfnamefont {Y.}~\bibnamefont
  {Nakata}},\ and\ \bibinfo {author} {\bibfnamefont {D.}~\bibnamefont {Reeb}},\
  }\bibfield  {title} {\bibinfo {title} {Fundamental energy cost for quantum
  measurement},\ }\href {http://arxiv.org/abs/1609.06981} {\bibfield  {journal}
  {\bibinfo  {journal} {arXiv:1609.06981 [cond-mat, physics:quant-ph]}\ }
  (\bibinfo {year} {2018})}\BibitemShut {NoStop}%
\bibitem [{\citenamefont {Buffoni}\ \emph {et~al.}(2019)\citenamefont
  {Buffoni}, \citenamefont {Solfanelli}, \citenamefont {Verrucchi},
  \citenamefont {Cuccoli},\ and\ \citenamefont
  {Campisi}}]{buffoni_quantum_2019}%
  \BibitemOpen
  \bibfield  {author} {\bibinfo {author} {\bibfnamefont {L.}~\bibnamefont
  {Buffoni}}, \bibinfo {author} {\bibfnamefont {A.}~\bibnamefont {Solfanelli}},
  \bibinfo {author} {\bibfnamefont {P.}~\bibnamefont {Verrucchi}}, \bibinfo
  {author} {\bibfnamefont {A.}~\bibnamefont {Cuccoli}},\ and\ \bibinfo {author}
  {\bibfnamefont {M.}~\bibnamefont {Campisi}},\ }\bibfield  {title} {\bibinfo
  {title} {Quantum {Measurement} {Cooling}},\ }\href
  {https://doi.org/10.1103/PhysRevLett.122.070603} {\bibfield  {journal}
  {\bibinfo  {journal} {Physical Review Letters}\ }\textbf {\bibinfo {volume}
  {122}},\ \bibinfo {pages} {070603} (\bibinfo {year} {2019})}\BibitemShut
  {NoStop}%
\bibitem [{\citenamefont {Solfanelli}\ \emph {et~al.}(2019)\citenamefont
  {Solfanelli}, \citenamefont {Buffoni}, \citenamefont {Cuccoli},\ and\
  \citenamefont {Campisi}}]{solfanelli_maximal_2019}%
  \BibitemOpen
  \bibfield  {author} {\bibinfo {author} {\bibfnamefont {A.}~\bibnamefont
  {Solfanelli}}, \bibinfo {author} {\bibfnamefont {L.}~\bibnamefont {Buffoni}},
  \bibinfo {author} {\bibfnamefont {A.}~\bibnamefont {Cuccoli}},\ and\ \bibinfo
  {author} {\bibfnamefont {M.}~\bibnamefont {Campisi}},\ }\bibfield  {title}
  {\bibinfo {title} {Maximal energy extraction via quantum measurement},\
  }\href {https://doi.org/10.1088/1742-5468/ab3721} {\bibfield  {journal}
  {\bibinfo  {journal} {Journal of Statistical Mechanics: Theory and
  Experiment}\ }\textbf {\bibinfo {volume} {2019}},\ \bibinfo {pages} {094003}
  (\bibinfo {year} {2019})}\BibitemShut {NoStop}%
\bibitem [{\citenamefont {Seah}\ \emph {et~al.}(2020)\citenamefont {Seah},
  \citenamefont {Nimmrichter},\ and\ \citenamefont
  {Scarani}}]{seah_maxwells_2020}%
  \BibitemOpen
  \bibfield  {author} {\bibinfo {author} {\bibfnamefont {S.}~\bibnamefont
  {Seah}}, \bibinfo {author} {\bibfnamefont {S.}~\bibnamefont {Nimmrichter}},\
  and\ \bibinfo {author} {\bibfnamefont {V.}~\bibnamefont {Scarani}},\
  }\bibfield  {title} {\bibinfo {title} {Maxwell's {Lesser} {Demon}: {A}
  {Quantum} {Engine} {Driven} by {Pointer} {Measurements}},\ }\href
  {https://doi.org/10.1103/PhysRevLett.124.100603} {\bibfield  {journal}
  {\bibinfo  {journal} {Physical Review Letters}\ }\textbf {\bibinfo {volume}
  {124}},\ \bibinfo {pages} {100603} (\bibinfo {year} {2020})}\BibitemShut
  {NoStop}%
\bibitem [{\citenamefont {Bresque}\ \emph {et~al.}(2021)\citenamefont
  {Bresque}, \citenamefont {Camati}, \citenamefont {Rogers}, \citenamefont
  {Murch}, \citenamefont {Jordan},\ and\ \citenamefont
  {Auffèves}}]{bresque_two-qubit_2021}%
  \BibitemOpen
  \bibfield  {author} {\bibinfo {author} {\bibfnamefont {L.}~\bibnamefont
  {Bresque}}, \bibinfo {author} {\bibfnamefont {P.~A.}\ \bibnamefont {Camati}},
  \bibinfo {author} {\bibfnamefont {S.}~\bibnamefont {Rogers}}, \bibinfo
  {author} {\bibfnamefont {K.}~\bibnamefont {Murch}}, \bibinfo {author}
  {\bibfnamefont {A.~N.}\ \bibnamefont {Jordan}},\ and\ \bibinfo {author}
  {\bibfnamefont {A.}~\bibnamefont {Auffèves}},\ }\bibfield  {title} {\bibinfo
  {title} {Two-{Qubit} {Engine} {Fueled} by {Entanglement} and {Local}
  {Measurements}},\ }\href {https://doi.org/10.1103/PhysRevLett.126.120605}
  {\bibfield  {journal} {\bibinfo  {journal} {Physical Review Letters}\
  }\textbf {\bibinfo {volume} {126}},\ \bibinfo {pages} {120605} (\bibinfo
  {year} {2021})}\BibitemShut {NoStop}%
\bibitem [{\citenamefont {Manikandan}\ \emph {et~al.}(2022)\citenamefont
  {Manikandan}, \citenamefont {Elouard}, \citenamefont {Murch}, \citenamefont
  {Auff{\`{e}}ves},\ and\ \citenamefont
  {Jordan}}]{manikandan_efficiently_2021}%
  \BibitemOpen
  \bibfield  {author} {\bibinfo {author} {\bibfnamefont {S.~K.}\ \bibnamefont
  {Manikandan}}, \bibinfo {author} {\bibfnamefont {C.}~\bibnamefont {Elouard}},
  \bibinfo {author} {\bibfnamefont {K.~W.}\ \bibnamefont {Murch}}, \bibinfo
  {author} {\bibfnamefont {A.}~\bibnamefont {Auff{\`{e}}ves}},\ and\ \bibinfo
  {author} {\bibfnamefont {A.~N.}\ \bibnamefont {Jordan}},\ }\bibfield  {title}
  {\bibinfo {title} {{Efficiently fueling a quantum engine with incompatible
  measurements}},\ }\href {https://doi.org/10.1103/PhysRevE.105.044137}
  {\bibfield  {journal} {\bibinfo  {journal} {Physical Review E}\ }\textbf
  {\bibinfo {volume} {105}},\ \bibinfo {pages} {044137} (\bibinfo {year}
  {2022})}\BibitemShut {NoStop}%
\bibitem [{\citenamefont {Campaioli}\ \emph {et~al.}(2017)\citenamefont
  {Campaioli}, \citenamefont {Pollock}, \citenamefont {Binder}, \citenamefont
  {C{\'{e}}leri}, \citenamefont {Goold}, \citenamefont {Vinjanampathy},\ and\
  \citenamefont {Modi}}]{Campaioli2017}%
  \BibitemOpen
  \bibfield  {author} {\bibinfo {author} {\bibfnamefont {F.}~\bibnamefont
  {Campaioli}}, \bibinfo {author} {\bibfnamefont {F.~A.}\ \bibnamefont
  {Pollock}}, \bibinfo {author} {\bibfnamefont {F.~C.}\ \bibnamefont {Binder}},
  \bibinfo {author} {\bibfnamefont {L.}~\bibnamefont {C{\'{e}}leri}}, \bibinfo
  {author} {\bibfnamefont {J.}~\bibnamefont {Goold}}, \bibinfo {author}
  {\bibfnamefont {S.}~\bibnamefont {Vinjanampathy}},\ and\ \bibinfo {author}
  {\bibfnamefont {K.}~\bibnamefont {Modi}},\ }\bibfield  {title} {\bibinfo
  {title} {{Enhancing the Charging Power of Quantum Batteries}},\ }\href
  {https://doi.org/10.1103/PhysRevLett.118.150601} {\bibfield  {journal}
  {\bibinfo  {journal} {Physical Review Letters}\ }\textbf {\bibinfo {volume}
  {118}},\ \bibinfo {pages} {150601} (\bibinfo {year} {2017})}\BibitemShut
  {NoStop}%
\bibitem [{\citenamefont {Ferraro}\ \emph {et~al.}(2018)\citenamefont
  {Ferraro}, \citenamefont {Campisi}, \citenamefont {Andolina}, \citenamefont
  {Pellegrini},\ and\ \citenamefont {Polini}}]{Ferraro2018}%
  \BibitemOpen
  \bibfield  {author} {\bibinfo {author} {\bibfnamefont {D.}~\bibnamefont
  {Ferraro}}, \bibinfo {author} {\bibfnamefont {M.}~\bibnamefont {Campisi}},
  \bibinfo {author} {\bibfnamefont {G.~M.}\ \bibnamefont {Andolina}}, \bibinfo
  {author} {\bibfnamefont {V.}~\bibnamefont {Pellegrini}},\ and\ \bibinfo
  {author} {\bibfnamefont {M.}~\bibnamefont {Polini}},\ }\bibfield  {title}
  {\bibinfo {title} {{High-Power Collective Charging of a Solid-State Quantum
  Battery}},\ }\href {https://doi.org/10.1103/PhysRevLett.120.117702}
  {\bibfield  {journal} {\bibinfo  {journal} {Physical Review Letters}\
  }\textbf {\bibinfo {volume} {120}},\ \bibinfo {pages} {117702} (\bibinfo
  {year} {2018})}\BibitemShut {NoStop}%
\bibitem [{\citenamefont {Binder}\ \emph {et~al.}(2015)\citenamefont {Binder},
  \citenamefont {Vinjanampathy}, \citenamefont {Modi},\ and\ \citenamefont
  {Goold}}]{Binder2015}%
  \BibitemOpen
  \bibfield  {author} {\bibinfo {author} {\bibfnamefont {F.~C.}\ \bibnamefont
  {Binder}}, \bibinfo {author} {\bibfnamefont {S.}~\bibnamefont
  {Vinjanampathy}}, \bibinfo {author} {\bibfnamefont {K.}~\bibnamefont
  {Modi}},\ and\ \bibinfo {author} {\bibfnamefont {J.}~\bibnamefont {Goold}},\
  }\bibfield  {title} {\bibinfo {title} {{Quantacell: powerful charging of
  quantum batteries}},\ }\href {https://doi.org/10.1088/1367-2630/17/7/075015}
  {\bibfield  {journal} {\bibinfo  {journal} {New Journal of Physics}\ }\textbf
  {\bibinfo {volume} {17}},\ \bibinfo {pages} {075015} (\bibinfo {year}
  {2015})}\BibitemShut {NoStop}%
\bibitem [{\citenamefont {Andolina}\ \emph {et~al.}(2019)\citenamefont
  {Andolina}, \citenamefont {Keck}, \citenamefont {Mari}, \citenamefont
  {Campisi}, \citenamefont {Giovannetti},\ and\ \citenamefont
  {Polini}}]{Andolina2019}%
  \BibitemOpen
  \bibfield  {author} {\bibinfo {author} {\bibfnamefont {G.~M.}\ \bibnamefont
  {Andolina}}, \bibinfo {author} {\bibfnamefont {M.}~\bibnamefont {Keck}},
  \bibinfo {author} {\bibfnamefont {A.}~\bibnamefont {Mari}}, \bibinfo {author}
  {\bibfnamefont {M.}~\bibnamefont {Campisi}}, \bibinfo {author} {\bibfnamefont
  {V.}~\bibnamefont {Giovannetti}},\ and\ \bibinfo {author} {\bibfnamefont
  {M.}~\bibnamefont {Polini}},\ }\bibfield  {title} {\bibinfo {title}
  {{Extractable Work, the Role of Correlations, and Asymptotic Freedom in
  Quantum Batteries}},\ }\href {https://doi.org/10.1103/PhysRevLett.122.047702}
  {\bibfield  {journal} {\bibinfo  {journal} {Physical Review Letters}\
  }\textbf {\bibinfo {volume} {122}},\ \bibinfo {pages} {047702} (\bibinfo
  {year} {2019})}\BibitemShut {NoStop}%
\bibitem [{\citenamefont {Juli{\`{a}}-Farr{\'{e}}}\ \emph
  {et~al.}(2020)\citenamefont {Juli{\`{a}}-Farr{\'{e}}}, \citenamefont
  {Salamon}, \citenamefont {Riera}, \citenamefont {Bera},\ and\ \citenamefont
  {Lewenstein}}]{Julia-Farre2020}%
  \BibitemOpen
  \bibfield  {author} {\bibinfo {author} {\bibfnamefont {S.}~\bibnamefont
  {Juli{\`{a}}-Farr{\'{e}}}}, \bibinfo {author} {\bibfnamefont
  {T.}~\bibnamefont {Salamon}}, \bibinfo {author} {\bibfnamefont
  {A.}~\bibnamefont {Riera}}, \bibinfo {author} {\bibfnamefont {M.~N.}\
  \bibnamefont {Bera}},\ and\ \bibinfo {author} {\bibfnamefont
  {M.}~\bibnamefont {Lewenstein}},\ }\bibfield  {title} {\bibinfo {title}
  {{Bounds on the capacity and power of quantum batteries}},\ }\href
  {https://doi.org/10.1103/PhysRevResearch.2.023113} {\bibfield  {journal}
  {\bibinfo  {journal} {Physical Review Research}\ }\textbf {\bibinfo {volume}
  {2}},\ \bibinfo {pages} {023113} (\bibinfo {year} {2020})}\BibitemShut
  {NoStop}%
\bibitem [{\citenamefont {Caravelli}\ \emph {et~al.}(2021)\citenamefont
  {Caravelli}, \citenamefont {Yan}, \citenamefont {Garc{\'{i}}a-Pintos},\ and\
  \citenamefont {Hamma}}]{Caravelli2021}%
  \BibitemOpen
  \bibfield  {author} {\bibinfo {author} {\bibfnamefont {F.}~\bibnamefont
  {Caravelli}}, \bibinfo {author} {\bibfnamefont {B.}~\bibnamefont {Yan}},
  \bibinfo {author} {\bibfnamefont {L.~P.}\ \bibnamefont
  {Garc{\'{i}}a-Pintos}},\ and\ \bibinfo {author} {\bibfnamefont
  {A.}~\bibnamefont {Hamma}},\ }\bibfield  {title} {\bibinfo {title} {{Energy
  storage and coherence in closed and open quantum batteries}},\ }\href
  {https://doi.org/10.22331/q-2021-07-15-505} {\bibfield  {journal} {\bibinfo
  {journal} {Quantum}\ }\textbf {\bibinfo {volume} {5}},\ \bibinfo {pages}
  {505} (\bibinfo {year} {2021})}\BibitemShut {NoStop}%
\bibitem [{\citenamefont {Tirone}\ \emph {et~al.}(2021)\citenamefont {Tirone},
  \citenamefont {Salvia},\ and\ \citenamefont {Giovannetti}}]{Tirone2021}%
  \BibitemOpen
  \bibfield  {author} {\bibinfo {author} {\bibfnamefont {S.}~\bibnamefont
  {Tirone}}, \bibinfo {author} {\bibfnamefont {R.}~\bibnamefont {Salvia}},\
  and\ \bibinfo {author} {\bibfnamefont {V.}~\bibnamefont {Giovannetti}},\
  }\bibfield  {title} {\bibinfo {title} {{Quantum Energy Lines and the Optimal
  Output Ergotropy Problem}},\ }\href
  {https://doi.org/10.1103/PhysRevLett.127.210601} {\bibfield  {journal}
  {\bibinfo  {journal} {Physical Review Letters}\ }\textbf {\bibinfo {volume}
  {127}},\ \bibinfo {pages} {210601} (\bibinfo {year} {2021})}\BibitemShut
  {NoStop}%
\end{thebibliography}
\end{document}